\documentclass[sigconf]{acmart}
\makeatletter                   
\def\mdseries@tt{m}             
\makeatother                    
\usepackage[draft=true]{minted} 
\usepackage{color}
\usepackage{hyperref}           
\hypersetup{
    colorlinks=true,
    linkcolor=blue,
    filecolor=red,      
    urlcolor=magenta,
    breaklinks=true,            
}
\usepackage{breakurl}           
\AtBeginDocument{%
  }

\usepackage[noend]{algpseudocode}
\usepackage[T1]{fontenc}
\usepackage[utf8]{inputenc}
\usepackage{algorithmicx}
\usepackage{algorithm}
\usepackage{amsmath}
\usepackage{array}
\usepackage{caption}
\usepackage{color}
\usepackage{enumitem}
\usepackage{float}
\usepackage{listings}
\usepackage{amsthm}
\usepackage{lstautogobble}
\usepackage{multirow}
\usepackage{nicefrac}
\usepackage{soul}
\usepackage{subcaption}
\usepackage{xcolor}
\usepackage{array}
\usepackage{thmtools}
\usepackage{tikz-qtree-compat}
\usepackage{tikz-qtree}
\usetikzlibrary{arrows.meta, positioning}
\usepackage{tikz}
\usepackage{listings}
\usepackage{cellspace}
\usepackage{latexsym}
\usepackage{supertabular}
\usepackage{tablefootnote}
\usepackage{longtable}
\usepackage{wrapfig}
\usepackage{refcount}
\usetikzlibrary{
    calc,
    positioning,
    er,
}
\usepackage{todonotes}
\usepackage{xspace}
\usepackage{alphabeta}
\newtheoremstyle{noIndentDef}  
  {3pt}   
  {3pt}   
  {\normalfont}  
  {}      
  {\bfseries} 
  {.}     
  { }     
  {\thmname{#1}~\thmnumber{#2} \thmnote{[#3]}} 

\theoremstyle{noIndentDef}

\setcopyright{none}
\renewcommand\footnotetextcopyrightpermission[1]{}

\definecolor{blue-cb}{RGB}{0,119,187}
\definecolor{cyan-cb}{RGB}{51,187,238}
\definecolor{teal-cb}{RGB}{0,153,136}
\definecolor{green-cb}{RGB}{34,136,51}
\definecolor{yellow-cb}{RGB}{170,156,49}
\definecolor{orange-cb}{RGB}{238,119,51}
\definecolor{red-cb}{RGB}{238,102,119}
\definecolor{magenta-cb}{RGB}{238,51,119}
\definecolor{purple-cb}{RGB}{170,51,119}
\definecolor{gray-cb}{RGB}{187,187,187}
\definecolor{c_mutable}{RGB}{32,167,178}

\newcolumntype{P}[1]{>{\centering\arraybackslash}p{#1}} 
\newcolumntype{L}[1]{>{\raggedright\let\newline\\\arraybackslash\hspace{0pt}}m{#1}}
\newcolumntype{R}[1]{>{\raggedleft\let\newline\\\arraybackslash\hspace{0pt}}m{#1}}

\newtheorem{definition}{Definition}
\newtheorem{example}{Example}
\newtheorem{corollary}{Corollary}
      
\AtBeginDocument{

}

\lstdefinelanguage{sql}{
    sensitive=false,
    keywords={select, from, where, as, in, and, distinct, create, drop, materialized, view, group, by, array_agg,
              left, outer, join, on},
    keywordstyle=\color{green-cb}, 
    morekeywords=[2]{resultdb},
    keywordstyle = {[2]\color{red-cb}},
    morekeywords=[3]{begin, transaction, commit},
    keywordstyle = {[3]\color{purple-cb}},
    string=[b]',
    morestring=[b]",
    stringstyle=\color{orange-cb},
    comment=[l]{--},
    morecomment=[s]{/*}{*/},
    commentstyle=\color{blue-cb},
}

\newfloat{algfloat}{htbp}{lop}
\floatname{algfloat}{Algorithm}

\algrenewcommand{\algorithmiccomment}[1]{\hfill\textcolor{blue-cb}{$\triangleright$ \textit{#1}}}

\AtBeginEnvironment{algorithm}{\small}
\DeclareCaptionFormat{algorithm}{\small\textbf{#1}#2#3}
\usetikzlibrary{
    arrows,
    calc,
    chains,
    matrix,
    positioning,
    scopes,
}
\usepackage{fancyvrb}

\begin{document}
\pagestyle{plain}
\title{How to get Rid of SQL, Relational Algebra, the Relational Model, ERM, and ORMs in a Single Paper --- A Thought Experiment}


\author{Jens Dittrich}
\affiliation{%
  \institution{Saarland University\\Saarland Informatics Campus}
  \city{Saarbrücken}
  \city{Saarland}
  \country{Germany}
}
\email{jens.dittrich@bigdata.uni-saarland.de}

\renewcommand{\shortauthors}{Dittrich}

\begin{abstract}

Without any doubt, the relational paradigm has been a huge success. At the same time, we believe that the time is ripe to rethink how database systems could look like if we designed them from scratch. Would we really end up with the same abstractions and techniques that are prevalent today? This paper explores that space. 

We discuss the various issues with both the relational model~(RM) and the entity-relationship model~(ERM). We provide a unified data model: the relational map type model (RMTM) which can represent both RM and ERM as special cases and overcomes all of their problems. 
We proceed to identify seven rules that an RMTM query language (QL) must fulfill and provide a foundation of a language fulfilling all seven rules. 
Our QL operates on \textit{maps} which may represent tuples, relations, databases or sets of databases. Like that we dramatically expand the existing operational abstractions found in SQL and relational algebra (RA) which only operate on relations/tables. In fact, RA is just a special case of our much more generic approach. 

This work has far-reaching consequences: we show a path how to come up with a modern QL that solves (almost if not) all problems of SQL. Our QL is much more expressive than SQL and integrates smoothly into existing programming languages (PL). 
In fact: you will have a hard time drawing a boundary between PL and QL. 
In our approach both QL and PL  become the `same thing', thus opening up some interesting holistic optimization opportunities between compilers and databases. In our QL, we also do not need to force application developers to switch to unfamiliar programming paradigms (like SQL or datalog): developers can stick with the abstractions provided by their PL.
Moreover, our QL makes SQL injection (as of 2025 still the 3rd most dangerous software weakness worldwide) impossible --- by definition. In addition, our query language fixes the broken snapshot semantics prevalent in ORMs.

We also show results of an initial experiment showcasing that just by switching to our data model, and without changing the underlying query processing algorithms, we can achieve speed-ups of up to a factor~3.

We will conclude that, if we build a database system from scratch, we could and should do this without SQL, RA, RM, ERM, and ORMs.
\end{abstract}

\maketitle

\section{Introduction}

Dear reader, you just read the title and the abstract of this paper and you are already getting mad and emotional about it? Fantastic! For your information: the  author of this paper is a professor of databases with more than 15 years experience in teaching undergrad and graduate database courses, a lover of relational technology, and a lover of the abstractions we are using --- be it to model data or to optimize queries. 

But then: every time I teach these abstractions, there are these friction points: why is X defined like this? Does Y really have to be like this? But why? Why is Z so complicated? Why do people need to work around Z then? In fact, are the abstractions we are using really the best possible abstractions we can come up with? In other words: \textit{If our existing abstractions and techniques are perfect, then why are there all these friction points?}

This paper is a thought experiment. I question the very foundations of relational database technology. Not necessarily to discard them entirely (we will see) but to come up with something better, simpler, more expressive, more versatile, and more universal. 

So, Dear reader, if you are still upset, I completely understand that feeling. I am with you. Relax, take a seat. What about letting some crazy thoughts slip into your mind? Enjoy the ride!

This paper is inspired by two very recent papers which we believe give a new spin to what databases can and should be.
The first, published at CIDR in January 2025~\cite{Deshpande} (best paper award), is a vision paper that proposes to keep the entity relationship abstraction as a DDL interface to the DBMS. So rather than \textit{first} translating an ERM to an RM to \textit{then} \texttt{CREATE} tables, that paper allows a DBMS to work with the ERM abstraction \textit{directly}.
This has many positive implications including: all semantics of the ERM are preserved. In addition, we obtain \textit{RM independence}: we can automatically create an optimal RM rather than hand-coding it. 
However, in terms of query language, \cite{Deshpande} is sketchy and suggests to resort to approaches like SQL++\cite{ong_sql++} which however leads to a couple of additional problems including the impedance mismatch with JSON.

A second recent paper, appearing in June 2025 at SIGMOD~\cite{Nix2025ExtendingSQL}, identifies a long list of problems with SQL which all have the same root cause: SQL forces all result data into a single (possibly denormalized) result relation. Therefore the authors suggest to extend SQL to allow it to return a result \textit{subdatabase}, i.e.~the relations with their subset of tuples from all the input relations of the query that contribute to the result. Those results are not shoehorned into a single output stream, but are returned as separate streams.

Those two works heavily inspired our paper. Originally, we planned to simply combine those two ideas. But while doing so, we observed that we can come up with a much more versatile data model and query language. Therefore, our paper is more than the  sum of~\cite{Deshpande} and~\cite{Nix2025ExtendingSQL}. It is the product.
This paper is structured as follows: in contrast to the common \textit{problem$\rightarrow$solution}-writing pattern, in Section~\ref{sec:RMTMpart1}, we will already start with parts of our solution. Concretely, this paper is structured as follows: 

\hspace*{-.35cm}
\begin{tikzpicture}[
  node distance=.4cm and 1.5cm,
  every node/.style={align=center, inner sep=2pt, outer sep=0pt, fill=black!15, minimum height=0pt},
  box/.style={draw=none},
  arrow/.style={->, thick, blue},
  scale=0.7, transform shape
]

\node (start) [box] {};
\node (RMT) [box, right=of start, xshift=-0.9cm] {introduce Maps \& Map Types\\(RMT, \autoref{sec:RMTMpart1})};
\node (RM) [box, right=of RMT, xshift=-.6cm] {redefine Relational Model\\(RM, \autoref{sec:RM})};
\node (RMC) [box, right=of RM, xshift=-0.5cm] {criticize RM\\(\autoref{sec:RMcriticism})};
\node (ERM) [box, below=of RM] {redefine Entity-Relationship Model\\(ERM, \autoref{sec:ERM})};
\node (ERMC) [box, right=of ERM, xshift=-0.7cm] {criticize ERM\\(\autoref{sec:ERMcriticism})};
\node (RMTM) [box, below left=.3cm of ERM] {introduce Relational Map Type Model\\(RMTM, \autoref{sec:RMTM})};
\node (rules) [box, below=of RMTM] {identify 7 Rules for an RMTM\\Query Language\\(\autoref{sec:DBQL})};
\node (query) [box, below=of rules] {introduce Map Views\\(\autoref{sec:MapViews})};
\node (queryusecases) [box, right=of rules] {show expressive power of Map Views\\(\autoref{sec:threeclassesofEVs})};
\node (RW) [box, right=of query] {RW\\(\autoref{sec:RW})};
\node (experiments) [box, right=of RW, xshift=-0.9cm] {Experiments\\(\autoref{sec:experiments})};
\node (conclusions) [box, right=of experiments, xshift=-0.9cm] {Conclusions\\(\autoref{sec:gettingrid})};

\draw[arrow]  (start) -- (RMT.west);
\draw[arrow] (RMT.east) -- (RM.west);
\draw[arrow] (RM.south) -- (ERM.north);

\draw[arrow] (RMC.west) -- ++(-0.3,0) -- (RM.east);
\draw[arrow] (RM.east) -- ++(0.3,0) -- (RMC.west);

\draw[arrow] (ERMC.west) -- ++(-0.3,0) -- (ERM.east);
\draw[arrow] (ERM.east) -- ++(0.3,0) -- (ERMC.west);
\draw[arrow] (RMC.east) |- (RMTM);

\draw[arrow] (ERMC) |- (RMTM);
\draw[arrow] (query.east) -- (queryusecases.west);

\draw[arrow] (RMTM.south) -- (rules.north);
\draw[arrow] (rules.south) -- (query.north);

\draw[arrow] (queryusecases.south) -- (RW.north);
\draw[arrow] (RW.east) -- (experiments.west);
\draw[arrow] (experiments.east) -- (conclusions.west);

\end{tikzpicture}

\section{Maps and Map Types}

\label{sec:RMTMpart1}

To start we need a couple of basic definitions. These definitions may feel very basic, but you will see that for the upcoming discussions we have to be very precise here.

\begin{definition}[Instance]
An instance is any element from a domain $D$ (aka type).
\end{definition}

\begin{definition}[Key]
A key $K_i$  is an instance from a domain~$D_K$.
\end{definition}

\begin{definition}[Value]
A value $V_j$  is an instance from a domain~$D_V$.
\end{definition}

\begin{definition}[Assignment]
A mapping of a key/domain-pair $(K_i, D_K)$ to a value/domain-pair $(V_j, D_V)$ is called an assignment:
$(K_i, D_{K_i}): (V_j, D_{V_j})$.
\end{definition}

\begin{definition}[Exclusive Assignment]
\label{def:exclusive:value}
An assignment of a value/domain-pair $(V_j, D_V)$ is called exclusive if $V_j$ is not mapped to from to any other key $K_{i\neq j}$ .
\end{definition}

\begin{definition}[Non-Exclusive Assignment]
\label{def:nonexclusive}
An assignment of a value/domain-pair $(V_j, D_V)$ is called  non-exclusive if $V_j$ is potentially mapped to from any other key $K_{i\neq j}$ as well.
\end{definition}

\begin{definition}[Map]
\label{def:MappingType}
A collection of assignments is called a \textit{map}.
\[
map := \big\{ (K_1, D_{K_1}): (V_1, D_{V_1}), \ldots,  (K_n, D_{K_n}): (V_n, D_{V_n})\big\}.
\]
\end{definition}
\begin{definition}[Map Type (MT)]
\label{def:MappingType}
A Map Type (MT) constrains (aka types) a map.
There are five possible constraints (and combinations thereof):
\begin{enumerate}[leftmargin=1.5cm]
\item [(Cn)] constrain $n$, i.e., the number of key/value-mappings,
\item [(CKD)] constrain the key domains  $D_{K_i}$,
\item [(CVD)] constrain the value domains $D_{V_i}$,
\item [(CK)] constrain the keys, i.e.~the $K_i$ instances,
\item [(CV)] constrain the values, i.e.~the $V_i$ instances.
\end{enumerate}
\end{definition}

For instance, a simple CKD-constraint is $K_i=$ int and a CVD-constraint is $V_i=$ str, i.e., a map that maps keys of type int to values of type string. 
Note that a map type constrains the structure of a single map not a set of maps.

\begin{definition}[Relational Map Type (RMT or Tuple Type)]
\label{def:RelationalType}
A relational map type (RMT) is a Map Type with at least the following constraints:
\begin{enumerate}[leftmargin=1.5cm]
\item [(Cn)]  $n=c$, where $c>0$ is a constant,
\item [(CKD)]  $D_{K_i}=$ symbolic names domain,
\item [(CK)] all keys $K_i,\ldots,K_n$ are defined (aka the attribute names).
\end{enumerate}

\end{definition}

For instance, in our running example ERM in~\autoref{fig:runningexample}, for the entity type `Professors', a valid RMT is:

$n=3$, $K_1$=id, $K_2$=name, $K_3$=age.  

Here, the domain of all keys are from a symbolic variable name domain \textit{sm}. 
However, in principle, key domains may be changed arbitrarily\footnote{Compare this with JSON which only allows keys to be strings.}. For instance, in order to represent the result of a pivot transformations, we need to be able to use arbitrary data as `column names', i.e.~keys in a map.

For simplicity, in the following, in order to specify an RMT, we will only denote the constraints CK and CVD and write this as follows:
\[
RMT := \big\{ K_1: D_{V_1}, \ldots,  K_n: D_{V_n}\big\}.
\]

\begin{definition}[Optional and Mandatory Keys]
\label{def:RelationalTypeoptionalattributes}
In a map type, a subset of the keys in a CK-constraint may be marked as optional:
$
MT := \big\{  \ldots,  [K_i: D_{V_i}], \ldots \big\}.
$
Here, $K_i,$ is an optional key. Keys that are not marked with `$[]$' are called \textit{mandatory}.
\end{definition}
\noindent For instance, for our running example ERM in~\autoref{fig:runningexample}, we could define:
Professors2 : \{ id: int, name: str, age: int, [dob:str] \}.

\noindent This indicates, that we do not have to assign a value to  key `dob'.

\vspace*{0.7em}
\noindent Through optional keys we avoid SQL's infamous NULL-semantic~\cite{grant_null-values}. An optional key of a map that has no value assigned corresponds to a NULL-value in SQL\footnote{In the relational model this is not possible and can only be simulated through the 6th normal form~\cite{DDL02}: split the relation into separate relations, one for each column: each relation has two keys: the tuple id and the actual value.} and its counterparts in ORMs.

\begin{definition}[Higher-Order Map Type (HOMT)]
\label{def:HOMT}
A~Higher-Order Map Type (HOMT) is a map type where at least one of the key or value domains is a map type. 
\end{definition}

\begin{definition}[Order of a Map Type]
\label{def:order}
Given a map type MT, the function $\textsf{order}(MT)\mapsto [0,\dots, +\infty)$ defines the nested-ness (or \textit{order}) of an MT.
An~MT where none of its key or value domains is an MT has order 0. An MT with domains of order of at most 0 has order 1, and so forth, recursively.
\end{definition}

\begin{definition}[Relational HOMT (RHOMT, Relation Type)]
\label{def:RHOMT}
A relational HOMT~(RHOMT) is a HOMT where the value domain of all values is the same RMT: (CVD) $\forall_{i\in [0,\ldots,n]}. \;V_{D_i}=$ RMT.

\end{definition}
Thus, a HOMT  is a higher order map mapping $n$ keys to maps.

\begin{definition}[Relation Map (RelMap)]
\label{def:RelMap}
Every map constrained by an RHOMT is called a \textit{Relation Map (RelMap)}.  

\end{definition}
\noindent Example:

\begin{small}
	profs\_RelMap: RelMap[Professors] = \{ 	
	
	~~~~~~~~~~~~~~~a\_dude: \{  id: 42, name: "Luke", age: 46, dob: 21-6-1979 \}.

	~~~~~~~~~~~~~~~some\_dude: \{ id: 31, name: "Horst", age: 25 \}, 

	\} 	
\end{small}

\noindent Here, the key domains are again symbolic. Alternatively we could define keys to be a function of the value. For instance,
Let $V_i$ be a value. Then, the function $\pi(V_i)$ returns a key $K_i$, e.g.~this could be a subset of the key/value-assignments of $m$.
Like that we can define keys to be `computed', e.g.~if we define $\pi$ to return the value of key `id', we obtain:

\begin{small}
	profs\_RelMap: RelMap[Professors] = \{ 
		
	~~~~~~~~~~~~~~~42: \{  id: 42, name: "Luke", age: 46, dob: 21-6-1979 \}.

	~~~~~~~~~~~~~~~31: \{ id: 31, name: "Horst", age: 25 \}, 

	\} 	
\end{small}

Thus, with computed keys, we have a natural way to express (primary) keys from Codd's relational model. In this model the key is `outside' the actual data and not forced into the actual data model.
The story behind this is much bigger and we can use much more general key computation functions here, e.g.~the ones we will later introduce for query processing with~\autoref{def:DBView}. For space constraints, we leave a deeper discussion of keys to future work. 

Note that as all maps contained in a RelMap must have the same RMT, this is actually the reason why querying such data is easy.
And at the same time this explains why querying unstructured data, where the uniformity of the RMTs is not given, is difficult. 

Still, within the same RelMap, it may happen that some maps have some subset of their optional attributes assigned while other maps do not assign these attributes (see~\autoref{def:RelationalTypeoptionalattributes}). Thus, RelMaps allow for a certain amount of unstructured-ness.

\begin{definition}[Maps are Enumerations]
\label{def:mapenum}
The values that are assigned to keys in a map define an enumeration. Such an enumeration is a valid value domain.
\end{definition}

We leverage enumerations in key domains to link to values already assigned to keys in other maps non-exclusively (see~\autoref{def:nonexclusive}). Notice the difference over a much looser type-based definition: we do not say that values must be of a ``certain type'', and we also do not say that only particular instances are allowed. Instead, we say that a concrete map defines an enumeration. 

For instance, assume a value domain $D_V$. There are at least four cases to constrain $D_V$:
\begin{enumerate}[leftmargin=1.15cm]
\item [(Case~1)] $D_V$ = int; i.e.~only integers are allowed,
\item [(Case~2)] $D_V$ = int $\wedge V\in K_{pk}$ of some other map; i.e.~like (1.), in addition: the value must exist as a value of a `primary key' in some other map,
\item [(Case~3)] $D_V$ = RMT; i.e.~the value is \textbf{some} instance of that RMT, \textbf{any} map complying with that RMT, not a specific map instance,
\item [(Case~4)] $D_V$ = RelMap[RMT]; i.e.~the value is an instance of a concrete RelMap[RMT], the value is a \textbf{concrete} value used in RelMap[RMT]
\end{enumerate}
We believe that Case~4 is the most natural way to model `links' between maps.

\section{The Relational Model}
\label{sec:RM}

Based on the definitions from Section~\ref{sec:RMTMpart1}, now, we can easily \linebreak\relax (re-)define the relational model (RM) as follows:

\begin{definition}[Relational Variable]
\label{def:RV}
A relational variable~(RV) is a relational map type~(RMT) (\autoref{def:RelationalType}) where all variable value assignments (\autoref{def:exclusive:value}) are exclusive. 
All non-exclusive key assignments must be translated into `foreign keys', see (Case~2):
$
RV :=  RMT \text{ where }~\forall i. K_i \text{ is exclusive} .
$
\end{definition}

\noindent\fbox{\begin{minipage}{.975\columnwidth}
\textbf{Important:}

The only thing an RV (Definition~\ref{def:RV}) ``adds'' over RMTs (Definition~\ref{def:RelationalType}) is a restriction, i.e.~RVs reduce the modeling power of RMTs.
\end{minipage}}

\begin{definition}[Relation]
A \textit{relation} is a \textbf{set} of an RV type.
\end{definition}
Notice the important difference here: a relation is \textit{not} a map but a set. In contrast, a RelMap (see~\autoref{def:RelMap}) is still a map.

\section{A Criticism of the Textbook-Style Relational Model}
\label{sec:RMcriticism}

\noindent\textbf{Where is the data?} The entire notion of who owns an instance is very implicit in the relational model. RM secretly assumes that attributes are exclusive. Thus each `cell' of a table (the primary visualization used for a relation) is actually a cell in the sense of a \textit{cage} for that instance. Attributes do not point to anything outside that cell, the cell \textit{contains} that something. In other words, the major defining property of RM over a relational map type~(RMT) is in RM attributes must not share their attribute values --- be it inside an instance of an RV or across different RV instances.

\noindent\textbf{The relation's type and the instance assigned to a variable are the same thing.} In textbook style RM as well as in SQL we mix up the \textit{type} of a relation and the \textit{instance} of such relational map type assigned to a variable: 
In SQL, if we define a table using \texttt{CREATE}~\texttt{TABLE}, at the same time, we are defining the type \textit{and} creating a single instance of that type. Then, the name of the type is used synonymously for the name of the single instance of that type. Only in SQL 1999, user-defined types and the \texttt{CREATE} \texttt{TYPE} statement were introduced. There is a great discussion about this problem in Chris Date's book~\cite{date} who fixes this for the relational model by introducing two separate terms \textit{relvar} (for the type) and \textit{relation} (for an instance of that type).

\noindent\textbf{Implementing Non-Exclusive Attributes through Foreign Keys.} 
If we want to reference instances from multiple places, we need to introduce a modified relational variable $RV_i$. Then we refer to maps of $RV_i$ through a \textit{foreign key mechanism}: the definition of $RV_j$ is changed to refer the primary key of $RV_i$ as a foreign key\footnote{Note that $j$ and $i$ may be equal for recursive relationships.}. This also implies that the domain(s) of the primary key attribute(s) used in $RV_i$ determine(s) the domain(s) of the foreign key attribute(s) in $RV_j$. In both definitions the domains must be the same (or at least `compatible'). In any way, in RM, we are repeating the domain information.
Technically, what RM does is to provide a specific \textit{data-model level implementation (to not say workaround) of non-exclusive assignments} (\autoref{def:nonexclusive}). Conceptually, there are no non-exclusive assignments in RM, as the instances used for foreign key attributes must be repeated. 

\begin{figure}[t!]
    \centering
	\includegraphics[trim = 0mm 165mm 485mm 0mm, clip,width=.55\columnwidth,keepaspectratio,page=8]{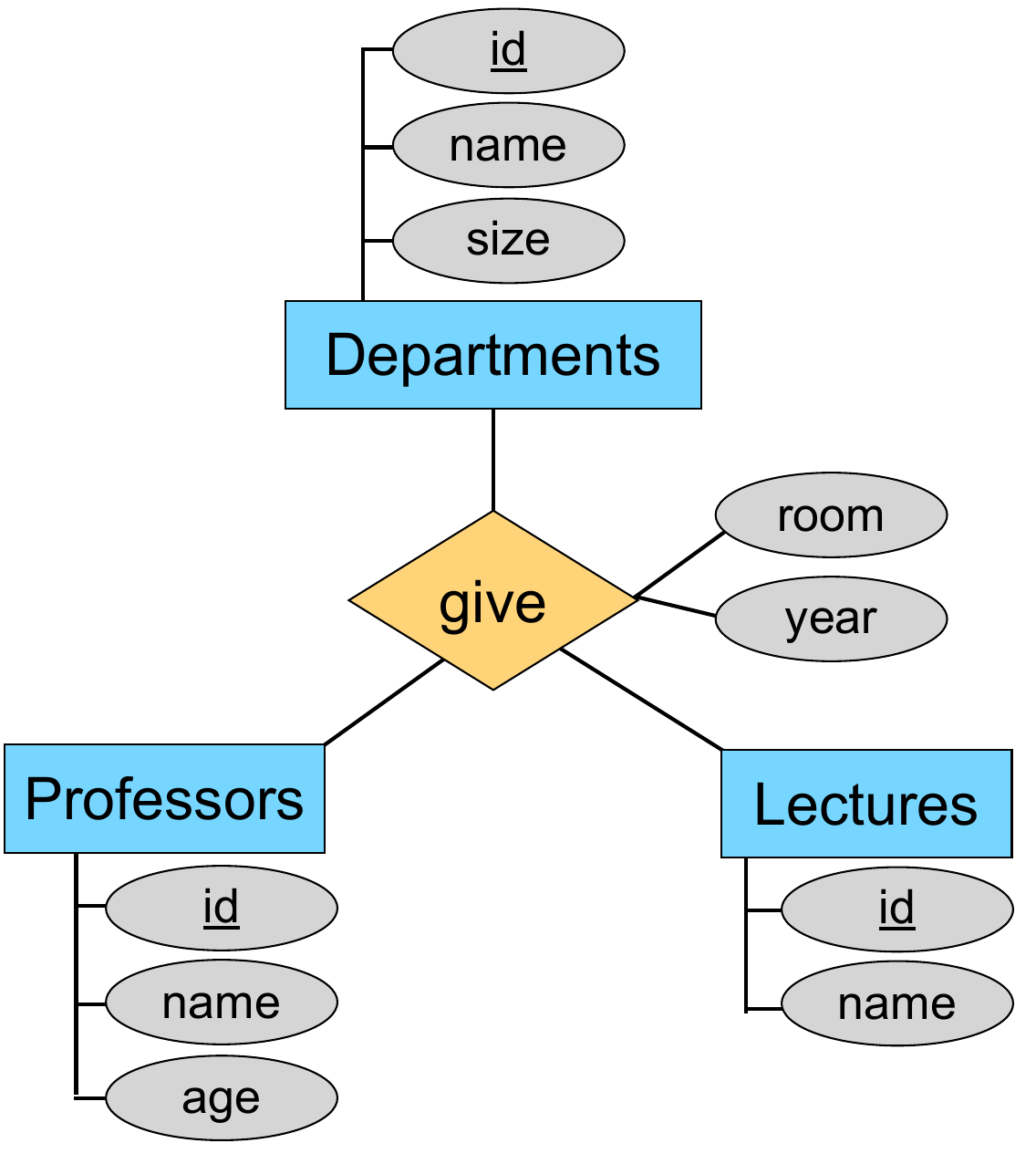}
\caption{Running Example\label{fig:runningexample} ERM}

\end{figure}

\noindent\textbf{Relations are Data Islands.} Each relation is like an independent data island with no real links/references to other relations. Links are created later by reconnecting and re\textit{interpreting} the relationships between foreign key and primary key attributes across tuples of relations through joins at query time. These join attribute values have to be kept in sync between foreign keys and primary keys. This is done by referential integrity constraints --- which is basically the RM version of avoiding dangling pointers. 

\noindent\textbf{Joins are Data Island-Connecting Operations.}  As RM does not allow for non-exclusive assignments we need to perform join operations in query processing. It is the analog of building bridges between the data islands --- at query time. And then we tear these bridges down again immediately after the query was processed (if we do not cache or materialize join results or and re-use the `intermediates' as workarounds). 

\noindent\textbf{Auto-ids.}  Most schemas we are aware of make heavy use of auto-ids. Alternatively, schemas use other types of artificial primary keys, typically some integer. Though there are situations where it may make sense to expose these keys beyond the realm of a DBMS, in many situations this is questionable. Why would you refer to an artificial key in your query if that key does not carry any semantics in your application anyways?
Querying artificial ids only makes sense if you use and expose them in one way or the other in your application: so you allow for queries like ``Hey did you see this awesome movie with auto-id 875913?''. Rather than phrasing a query like ``Hey did you see this awesome Luc Besson movie with Milla Jovovich and Bruce Willis?''  Again, we are not claiming that exposing (auto-)ids does never makes sense. But technically, we argue that (auto-)ids are internal data that should (and in most cases can) be hidden from the outside world.

\noindent\textbf{A Simple Fix for Auto-ids:}  Database systems offer various mechanisms to create auto ids. Here is the mechanism from PostgreSQL:
\begin{small}
\begin{lstlisting}[language=sql]
CREATE TABLE R (
    id INT PRIMARY KEY GENERATED ALWAYS AS IDENTITY,
    foo VARCHAR NOT NULL
);
\end{lstlisting}
\end{small}
This convoluted expressions says: we have an attribute named \textit{id} of type \texttt{int} which is the primary key of table \textit{R}. The attribute values are generated by the DBMS and serve as identity for the rows of this table.
Why not extend this to allow for the following:

\begin{small}
\begin{lstlisting}[language=sql]
CREATE TABLE R (
    IDENTITY HIDDEN,
    foo VARCHAR NOT NULL
);
\end{lstlisting}
\end{small}
Here, \texttt{HIDDEN}~\texttt{IDENTITY} could signal that the identity of rows is handled internally by the DBMS (in whatever ways). There is not even a name for the identity attribute, and we do not have to know about the attribute's domain used internally (if the DBMS is even using foreign keys internally, we actually do not care about this). And hence you won't be able to use `the identity attribute' in your queries.
Our proposal leads to a couple of reconsiderations for query optimization and processing. We will come back to this in Section~\ref{sec:qprocandopt} and in our experiments in Section~\ref{sec:exp:swizzlingvsjoins}.

\section{Entity-Relationship Model}
\label{sec:ERM}

In the previous section we discussed various issues with the relational model (RM). Now, let's do a similar discussion for the entity-relationship model (ERM).
Based on the definitions from Section~\ref{sec:RMTM}, we can easily (re-)define ERM as follows:

\begin{definition}[Entity Type (EntT)]
\label{def:EntT}
An entity type (EntT) is a relational variable (RV) where in addition to the constraints of RVs, the value domains \textbf{must neither} be Entity Types \textbf{nor} Relationship Types:
\begin{enumerate}[leftmargin=1.5cm]
\item [(CVD)] $\forall i. \text{ type}(D_{V_i})  \notin \{EntT,RelT\}.$
\end{enumerate}
\end{definition}
So, in fact EntTs and RVs are very similar. 

\begin{definition}[Relationship Type (RelT)]
\label{def:RelationshipType}
A relationship type~(RelT) is a relational variable~(RV) with at least two keys where in addition to the constraints of RVs, at least two of the value domains are entity types. In addition, none of the value domains is a RelT:

\begin{enumerate}[leftmargin=1.5cm]
\item [(Cn)]  $c>2$,
\item [(CVD)] $\forall i. \left|\{ \; i \;|\; \text{type}(D_{V_i})  = EntT \}\right| \geq 2$
\item [~~~ ~ ] $\wedge\;  \text{ type}(D_{V_i})  \ne RelT$
\end{enumerate}

\end{definition}

\noindent\fbox{\begin{minipage}{.975\columnwidth}
\textbf{Important:}

Notice that the only thing EntTs and RelTs (Definitions~\ref{def:EntT}\&\ref{def:RelationshipType}) ``adds'' over RMTs (Definition~\ref{def:RelationalType}) are restrictions, i.e.~EntTs and RelTs reduce the modeling power of RMTs even further.
\end{minipage}}

\begin{definition}[ER-model]
An ER-model (ERM) is a set of EntTs and RelTs.
\end{definition}

\begin{definition}[Entity Set]
An entity set ES of EntT is a \textbf{set} of maps of type EntT.
\end{definition}

\begin{definition}[Relationship Set]
A relationship set RS of RelT is a \textbf{set} of maps of type RelT.
\end{definition}

With these definitions, now, we can actually define an \textit{ER database}. The following definition formalizes which was recently proposed using a variant SQL DDL by Deshpande~\cite{Deshpande}:

\begin{definition}[ER Database]
An ER database (ERDB) consists of two sets: a set $\mathcal{ES}$ of all entity sets, and a set $\mathcal{RS}$ of all relationship sets. In addition, 
\[
\forall rs \in \mathcal{RS}, \ \forall m \in rs, \ \forall D_{V_i} \in rs: \left(D_{V_i} \neq RMT \vee D_{V_i} \in \mathcal{ES}\right).
\]
In other words, all value domains of all relationship set point to enumerations of entity sets present in the ER database, i.e.~no `link' points to something outside the database.
\label{def:ERDB}
\end{definition}

\section{A Criticism of Textbook-Style ERM}
\label{sec:ERMcriticism}

But why do EntTs and RelTs make these constraints?

\noindent\textbf{Structured Attributes.} Actually, some variants of ERM drop some of these constraints, e.g.~allow EntTs to have attributes of type EntT~\cite{ParentS84}. Like that they blurr the boundary between RelTs and EntTs even more.
This can also be observed each time we want to promote an atomic attribute, say of EntT$_1$, to have more structure.
In that case, we have to remove the attribute from EntT$_1$, add a new EntT$_2$ and connect EntT$_1$ and EntT$_2$ by an `intermediate' relationship type RelT. So, just by adding structure to an attribute, we use a completely different modeling construct.
This restriction of EntTs is probably due to the old idea of keeping relations in first normal form and restrict value domains to `atomic' types.
That `atomicity idea' along with the idea to keep data in first normal form was discarded long time ago in SQL 99: which allows for arbitrary relational entity types including arrays, JSON, and even nested tables.
\begin{figure}[t!]
    \centering
	\includegraphics[trim = 0mm 298mm 330mm 0mm, clip,width=\columnwidth,keepaspectratio,page=9]{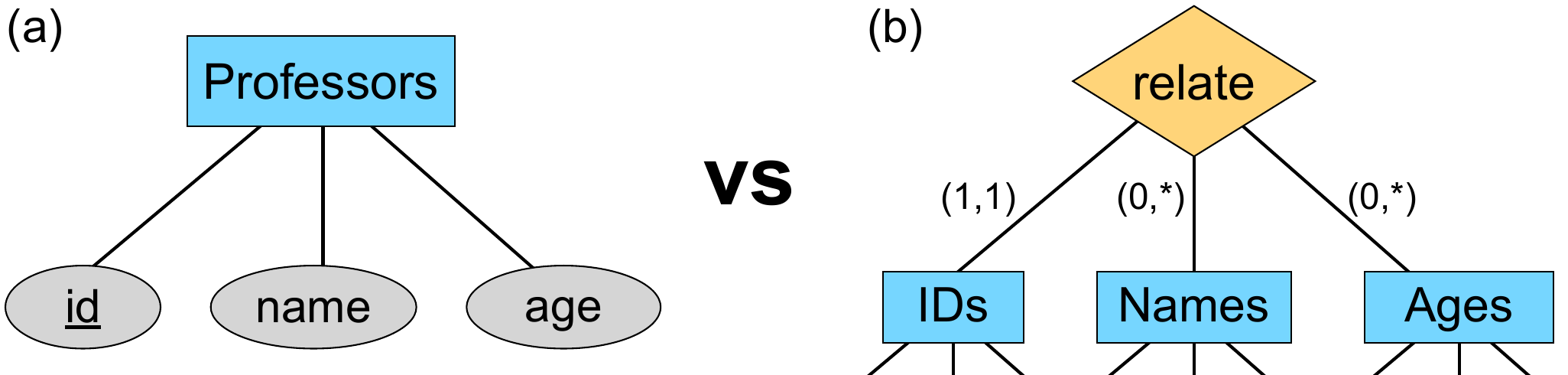}
    \caption{An entity types with attributes and an underlined key attribute vs a relationship type with three entity types and (min,max)-functionalities, one of them a (1,1)-functionality\label{fig:keyvsminmax}}
\end{figure}

\noindent\textbf{Different Modeling Constructs for the same Thing.} ERM introduces even more weirdnesses: even a single EntT like Professors connects its attributes in a relationship already. For instance, in our running example in Figure~\ref{fig:runningexample} entity type Professors has three attributes id, name, and age.
Any entity of Professors glues three attribute values together, in other words: \textbf{the EntT puts attributes values into a relationship}, in fact: \textit{any EntT already expresses a relationship among its attribute values}.
But, if we do the same thing one abstraction layer higher, i.e.~by combining three entities from EntTs Professors, Departments, and Lectures using an (`official') relationship type \textit{give}, only then we officially call it `a relationship'.
For this, we all of a sudden use different modeling constructs (rhombus vs rectangle), whereas mathematically we are doing \textbf{the exact same thing}: we are relating entities (or in the language of our model: `instances') from different value domains. 

\noindent\textbf{Keys on different levels.} This can also be seen if we remind ourselves about how keys are handled in ERM. Figure~\ref{fig:keyvsminmax} shows two different ways to model Professors: 
Figure~\ref{fig:keyvsminmax}(a) models Professors as an EntT. In contrast, Figure~\ref{fig:keyvsminmax}(b) models Professors as an RelT over different EntTs.
Those EntTs are promoted attributes with potentially additional attributes (not shown in the figure).
Again, whenever we want to promote an attribute to be structured, we need to promote that attribute to an EntT.

In summary, \textbf{just like EntTs, a RelT puts its attribute values into a relationship, just that the domains may be EntTs}.

So, why do we use two different modeling constructs to express the same thing? Again, this happens only because we distinguish between EntTs and RelTs in the first place!

\noindent\textbf{Translating ERMs to RMs.}
Let's make another journey into good old undergrad material. We teach our students to: First, design a proper ERM, and: Second, translate that ERM to RM.
This basically boils down to creating for each EntT and for each RelT a separate relation\footnote{We are aware that we are ignoring the shortcuts here: for some RelTs we do not have to create separate relations, e.g.~for 1:N and 1:1-relationships we simply add foreign key columns to one of the relations created for the participating EntTs and so forth.}.
In addition, we have to link the entities and relationships to each other.
For instance, if professor `Luke Skywalker' gives two lectures `Laser Swords' and `Mastering the Force', we add a relationship type to \textit{give}, \textbf{but:} we do \textbf{not} blindly include all that data from that professor map and all the data from the lectures, thus repeating the same information over and over again redundantly.
Instead, we use again the relational non-exclusive attribute `hack':

\noindent\textbf{Implementing References through Foreign Keys.} 
Instead, and as discussed already above for the relational model already, we use one specific implementation of non-exclusive attributes: in every relationship of relationship set `give' we add a reference: an attribute value that repeats the attribute value of the primary key of the participating entity set. We call this a \textit{foreign key}.
Like that, we do not have to repeat all data from `Luke Skywalker'  whenever he gives another lecture.
So, in summary, from the point of view of relationship set `give', entity set `professors' is a dictionary.

\noindent\textbf{Chasing References during Query Processing.} 
The price we pay for that is that, at query time we have to trace the relationships expressed by foreign keys. We achieve this by searching indexes and/or by performing joins. In other words, as we only have the foreign key-implementation of non-exclusive attributes, \textbf{at query time}, we have to connect the \st{dots} rows, to be able to concatenate individual tuples into larger tuples. 
In object-oriented databases this is done already while loading data from disk: references on disk are implemented as persistent IDs. Then, when the data is loaded into main memory, these IDs are swizzled to memory addresses.

\noindent\textbf{Relationships are always Weak!}
Traditionally, wen can mark entity types along with the relationship type to their strong `parent'  as weak, e.g.~a room belonging to a building disappears if you destroy the building. 
Weak entities cannot exist without the participating strong `parent' entity they reference.
However, conceptually, \textbf{any type of relationship is always `weak' by definition} as any relationship pointing to an entity that does not exist simply does not make sense. A relationship \textbf{always} depends on its participating entities. 
And the same is true for any attribute with a entity type-domain: if you manage to set a reference to refer to an entity that does not exist, i.e.~either by using a foreign key that does not exists or by setting to NULL), that conceptual relationship disappears into thin air.

So, in summary, we can differentiate between two types of `weaknesses': (1.)~\textbf{classical weak entity types} and (2.)~\textbf{weak relationship types}: relationship types are weak by definition as you simply cannot relate non-existing entities.
The latter would break the mathematical definition of a relation.

\section{A Simpler Model: The Relational Map Type Model~(RMTM)}
\label{sec:RMTM}

\newcounter{defDiff}
\setcounter{defDiff}{\getrefnumber{def:ERDB}}
\addtocounter{defDiff}{-\getrefnumber{def:RV}+1}

To get rid of all of these issues present in RM and ERM, there is a simple solution: let's drop the constraints imposed by RVs (used in RM), as well as the constraints of EntTs and RelTs (used in ERM) and simply use RMTs! In other words: get rid of the \arabic{defDiff} definitions~\ref{def:RV}--\ref{def:ERDB}.
Like that we can define:

\begin{definition}[Database Schema (DBS) aka Relational Map Type Model (RMTM)]
\label{def:DBS}
A database schema is a HOMT (\autoref{def:HOMT}) mapping keys to RMTs: (CVD)~$D_{V_i}= RMT_i$.
\end{definition}
For instance, assume we define an RMT:

\begin{small}
\[
\text{ProfsRMT := \{ id: int, name: str, age: int \}.}
\]
\end{small}
Then, we cans define a database schema as:
\begin{small}
\[
\text{My\_DBS := \big\{old\_profs: ProfsRMT, new\_profs: ProfsRMT\big\}.}
\]
\end{small}
Thus, we use the same  RMT twice in this schema. This allows us to keep the profs in two separate sets, e.g.~to partition those sets horizontally.

\noindent For our running example (\autoref{fig:runningexample}), a valid database schema is:

\begin{small}
\vspace*{0.7em}
running\_example\_DBS = \{

	~~~~~Professors: ProfsRMT,
	
	~~~~~Lectures:  \{ name: str \},
	
	~~~~~Departments:  \{ name: str,  size: int \},
	
	~~~~~give:  \{ p: Professors, l: Lectures, d: Departments \}
	
	~~~~~\}
\end{small}

\vspace*{0.7em}
Here, only the map Professors uses the predefined ProfsRMT.
The other RMT definitions are done in place in the database schema definition.
Notice the value domains used in RMT `give': the value domain of key p \textbf{is not} ProfsRMT, it is Professors, i.e.~the concrete map which also defines an enumeration (recall Definition~\ref{def:mapenum}).
Thus by using the enumeration (Case~4), we are signaling that the values of key p are maps (`tuples') existing in RelMap Professors, i.e.~we are linking (in the sense of a non-exclusive assignment, see~\autoref{def:nonexclusive}) to concrete maps in RelMap Professors, and not just some valid map constrained by ProfsRMT.

\begin{definition}[Database (DB)]
A database (\texttt{DB}) is a map constrained by a database schema. This means, for each $(K_i: RMT_i)$-entry contained in a DBS, a \texttt{DB} contains a map $m_i$ constrained by $RMT_i$:
$
\texttt{DB} := \big\{\; (K_i: map[RMT_i]) \;\;|\;\; (K_i: RMT_i) \in DBS \; \big\}.
$
\end{definition}

\noindent For instance, for our running example, a valid database is:

\begin{small}
\vspace*{0.7em}
running\_example\_DB = \{

	~~~~~Professors: \{~~42: \{ name: "Luke", age: 35 \}, 

	~~~~~~~~~~~~~~~~~~~~~~~~~~~31: \{ name: "Horst", age: 25 \} \}, 

	~~~~~~~~~~~~~~~~~~~~~~~~~~~32: \{ name: "Horst", age: 25 \} \}, 

	~~~~~Lectures:~~~ \{~~17: \{ name: "Databases are great"  \}, 

	~~~~~~~~~~~~~~~~~~~~~~~~~~~66: \{ name: "NOSQL sucks"  \} \}, .. \}
	
\end{small}

\vspace*{0.7em}
Keep in mind that this example is a \textit{conceptual notation}. The syntax we use does \textbf{not} imply that the keys (`attribute names') have to be physically repeated for each and every map (`tuple') like in JSON or XML. Also note that the identity of a key/value-entry of Professors is given by the key and not encoded in the value.

\begin{corollary}[Database are Enumerations]
\label{def:DBenum}
A DB is a map. Therefore, a DB is also an enumeration (recall~\autoref{def:mapenum}).
\end{corollary}

 And hence we can use a DB to express relationships not on a ``tuple''- but on a ``relation''-level. 
 \begin{figure}
\noindent\fbox{\begin{minipage}{.975\columnwidth}

\noindent\textbf{(1.)~No Domain Restrictions:}  in RMTM \textbf{we do not restrict the value domains in any way}. In particular, we do not assume references to maps or relationships to be implemented by foreign keys as in RM.

\noindent\textbf{(2.)~Relational Inception:} Translating these definitions back to standard relational terminology, this implies that the term `database' is just a fancy term/shorthand/alias for `a relation of relations', i.e.~a higher order relation.
A traditional relational database is nothing but a nesting of at least two levels where the the top-level relation (the `database') contains tuples containing the nested relations (the `relations' of the `database'). 
Pushing this further, a DBMS containing multiple databases is nothing but a nesting of at least three levels where the top-level relation contains tuples each representing an entire database. And so forth.

Also note. that if a `table' is allowed to contain `tables' as introduced in SQL 1999, we obtain more levels: a `table' with a `table' type attribute is nothing but what in RMTM we call a `database'. In SQL, it remains a `table' with `tuples' containing `tables'. Even though a tuple may contain tables, we do not call that tuple a database.

\textbf{Thus, in RMTM; everything is a map:} we use the same abstraction at all levels: \textit{maps}, i.e.~a key/value-mapping from keys (aka attributes) to `stuff'. In other words, in our approach \textbf{everything is a key-value-store or a higher-level (nested) version thereof}.

\begin{center}
\begin{tabular}{|l||l|}
\hline
\multicolumn{2}{|c|}{\textbf{Terminology}}  \\\hline\hline
\textbf{Relational Model} & \textbf{RMTM}  \\\hline\hline
tuple & map \\\hline
relation & map \\\hline
database & map \\\hline
set of databases & map \\\hline
\end{tabular}
\end{center}
This observation is not purely academic but has real-world implications: in our query language (see~Sections~\ref{sec:DBQL}~and~\ref{sec:MapViews}) we can treat entire databases as individual maps and use that e.g.~to split an entire database into multiple databases or vice versa\footnote{In this paper, for simplicity, we will stick with these four hierarchy-levels, even though more levels are possible.}.
\end{minipage}}
\caption{Differences of RM vs RMTM\label{fig:diffRMvsRMTM}}
\end{figure}

\subsection{Impact on Query Processing and Optimization}
\label{sec:qprocandopt}

As discussed in~\autoref{sec:RMcriticism}, the relational model creates separate data islands where each relation is a data island.
Foreign keys represent links between tuples.
And these links have to be reconnected at query time. For each query. Again and again and again.

An alternative would be to resolve these links \textit{only once}, e.g.~when the data gets loaded from disk. Object-oriented DBMSs do this and call it \textit{pointer swizzling}~\cite{KemperK93}. With RMTM, or in RM with our without using the \texttt{HIDDEN}~\texttt{IDENTITY} feature we propose, nobody stops us from doing a similar thing, also in a relational DBMS: 
When loading the data swap the external identity by a memory reference (vice versa when storing).
Alternatively, we could keep the memory references redundantly over the external identity as a separate column (in particular if the identity is visible outside the database).
This implies that for all data in main-memory all links through foreign keys get replaced by direct links. Thus, we can follow all links directly. 
And that changes query processing quite a bit.
Consider a standard star schema with a central fact table~$F$ and a couple of extra attributes serving as `measures', plus a couple of dimensions tables $D_1,\ldots,D_n$\footnote{Technically, even our running example in Figure~\ref{fig:runningexample} is a `fact table' if we interpret any of its attributes as a `measure'. We could go as far to argue that  any n-ary relationship with measures is indeed a `fact table'.}. 
The classical approach to compute such star join is to perform an n-ary join: 

\noindent\fbox{\begin{minipage}{.975\columnwidth}
\noindent\textbf{Star Join Algorithm (Relational Model):}

(1.)~Create (and keep) hash indexes on all primary keys of all dimension tables. 

(2.)~Loop over all tuples from $F$, for each tuple: lookup its foreign key in each hash table of the referred dimension table (as long as it matches).
\end{minipage}}

The rational behind this is that the fact table is typically very large and the dimensions tables are small. Thus, this join algorithm works well as long as all dimension tables fit into main memory. Think of it as the n-ary version of a simple hash join.
Additional improvements include join order, i.e.~try to probe the most selective hash table first.
You can view the same star join at the same time as conceptually performing an n-ary join (join $F$ at the same time with all dimension tables), or as a linear query execution plan (either left or right-deep) of binary joins (join $F$ with one of the dimensions first, then continue along the spine of the linear plan). 

In contrast, in our approach, each map of $F$ has direct links to the maps of the connected dimension tables $D_1,\ldots,D_n$.
Thus, we do not need any of these hash tables to identify matching rows which, depending on the organization of our store, then also have to be looked up in that store.

\noindent\fbox{\begin{minipage}{.975\columnwidth}
\noindent\textbf{Star Join Algorithm (Relational Map Type Model):}

\st{(1.)~Create (and keep) hash indexes on all primary keys of all dimension tables. }

(2.)~Loop over all tuples from $F$, for each tuple: lookup the map in the referred dimension table directly.
\end{minipage}}

So, in summary, in RMTM, we save one hop over the hash table. Depending on the size of the fact table and the number of dimensions tables involved, the cost savings may be substantial. We will evaluate this in our Experiments in Section~\ref{sec:exp:swizzlingvsjoins}.

The discussion of which tuples we still need to join at query time and which not definitely justifies a separate study. We leave this for future work.  Yet, there is another important aspect to that.

\subsection{But what does a Query on a Map `Return'?}
\label{sec:queryreturntype}

This is crucial to understand. Not understanding the implications of this question is the root cause of a lot of (performance) pain with databases.
An infamous real-world instance of this problem is the n+1 query problem.
It occurs a lot in ORMs, but you can also run into this trap through an unlucky interaction of application code and SQL-queries. 

Let's take our running example in Figure~\ref{fig:runningexample} and a textbook translation to RM. Now, in our application code, we select all  \textit{give} relationships that happened in year 2024 through an ORM query:
\begin{lstlisting}[language=Python]
results = give.objects.filter(year=2024)
for r in results:
   print(r.professor.name)
\end{lstlisting}
In this example, line~1 contains the actual query. Then, for each iteration in line~3, in order to resolve the value of \texttt{name}, the ORM will trigger one extra SQL-query to the \textit{Professors} relation. In other words, we trigger one additional SQL-query per iteration over the result set \texttt{results}. Hence the name `N+1' query problem. 

Obviously, no one with even minimal knowledge in relational database technology would/should do it like that.
The right way of doing this is to send one bigger query joining \textit{give} and \textit{Professors} and adding \textit{Professors.name} to the \texttt{SELECT}-clause.
Sure, but this happens nevertheless in ORMs.
One of the main purposes of ORMs\footnote{In terms of design patterns a better name for ORMs would actually be ORW: object-relational wrapper.} is to hide RM and SQL under an object-oriented façade.
In that world, like in RMTM, the application developer can navigate all links directly.
For Django ORM, you can easily inspect and profile this problem using tools like Django Debug Toolbar.
Django ORM offers  \textbf{workarounds for this problem} (e.g., \texttt{select\_related()}, \texttt{prefetch\_related()}) which allow you to declare upfront which attributes you will access later on and thus instruct the ORM to fetch those attribute values already with the initial query.
Unfortunately, in a complex code base, it is easy to overlook some of those attribute accesses. Therefore, there is a long list of Django extensions trying to tackle the problem by prefetching attributes (e.g., django-auto-prefetch~\cite{djangoautoprefetch}, django-seal~\cite{djangoseal}, and nplusone~\cite{djangonplusone}).

 In a purely relational world,  the N+1 problem simply does not exist (cf.~Definition~\ref{def:RV}: ``all attribute assignments are exclusive'') under the assumption that all attribute values are fetched directly with the query. However, this problem shows up as soon as we start marshaling RM into an object-oriented model to mimic non-exclusive assignments, as in Django ORM or any other ORM. And the N+1 problem may occur for our RMTM as well.

Also note the \textbf{misleading transactional semantics}  that come with the N+1 query problem: if you send one query and prefetch all attribute values you ever plan to access, you get a consistent view (aka snapshot) of the data. In contrast, if you first send a query, and then lookup other attributes afterwards outside the snapshot of that initial query, you might get an inconsistent view.
In fact, in the extreme case, each individual lookup of \texttt{name} may point to a different database version. In addition, some of your initial query \texttt{results} might already have been deleted, others might have been added. You may even create outputs violating your initial filter conditions, e.g.~if the \textit{year} attribute of the \textit{give}-relationships are concurrently modified to be equal or not equal to 2024.

To avoid this, you must wrap all your code into transactions \textbf{and} make sure to set the possibly weaker default isolation level (like \texttt{read committed} in PostgreSQL) to \texttt{serializable}. Otherwise you are stitching together data from possible different versions with a good chance of creating isolation anomalies (assuming that based on these reads you write something to the database\footnote{``...based on these reads you write something to the database.''. This does not even have to happen in a transaction to create an anomaly from the point of view of the application's or user's notion of consistency. Even if you do the reads in one transaction, then show the result to the user, and then based on that information the user issues a write in a different transaction, you may ruin consistency, as that read-write conflict is invisible to the DBMS. However, this problem is orthogonal to the data model and query language used.}). And you cannot blame the DBMS for these anomalies as this is neither under the control nor under the responsibility of the DBMS, but solely your own fault --- though you might have overlooked your responsibility for this.
The problem here is the \textbf{user's perception}: ``hey, I got my query results, now, let's grab a few more attribute values here and there and do some meaningful stuff with it.''
This is all fine, as long as all of this is running in a transaction; but if not, anything may happen and ruin the consistency of the DB.

Getting back to our initial question ``What does a query on a DB return?''. In a SQL-world of disconnected data islands the answer is clear: SQL returns a data island.
In contrast,  in a highly connected RMTM DB, the answer is more complex. This is because starting from any map in a DB you can reach each any other map of the DB anyways by following references (or back-references). So, what is the point of reducing the output of the query to a `data island', i.e.~a disconnected subset of the input DB?
We will see in the following section how this observation influences our DB query language.

\section{Seven~Rules Towards an RMTM QL} 
 \label{sec:DBQL}

An RMTM query language (QL) should adhere to the following seven rules:

\noindent\textbf{(1.)~The QL is an algebra where the input is a map and the output is a map}. As outlined in~\autoref{sec:RMTM}, in traditional database terminology a map may represent a tuple, a relation, a database, or a set of databases, etc. The QL should be able to operate on these different granularities.

\noindent\textbf{(2.)~The QL clearly separates between filtering and transforming data.} SQL/RA mixes up data filtering with the data transformation aspects. A QL should clearly splits the semantic of a query into two parts:

\begin{minipage}{.95\columnwidth}
\textbf{(A.) Identify} the relevant (sub-)maps from the input map. And optionally:

\textbf{(B.) Transform} that data into a desired output map.
\end{minipage}

\noindent Obviously, these steps may be combined and nested arbitrarily.

\noindent\textbf{(3.)~The QL is purely functional and non-textual and seamlessly blends into programming languages.} 
 Everything in the QL should be expressible through functions.  The functional nature of the QL should make it very easy to integrate it into existing PLs: the language should be views dressed up in a functions costume of the embedding PL.
 These function costumes can easily be provided by a PL~library allowing programmers to assemble expressions within any host PL, say \textit{MyPL}. 
However, these functions are \textbf{not necessarily} run inside MyPL but the entire expression or any suitable part of it \textbf{may be pushed down} to the database system which can then optimize the expression and return a map (through some streaming interface: ONC, generators, vectorized, etc.).
So from the point of view of the developer it looks like as if all those functions were executed in the PL --- and in the order specified. However, the DBMS may decide differently.
 Like that the QL can avoid the classical impedance mismatch of QL and PL types. In fact, the entire artificial boundary between SQL and the PL embedding SQL-expressions vanishes.

This also opens up interesting future work as now the optimizations done by the MyPL compiler/interpreter/runtime overlap with the optimizations done with the DBMS.
Thus, we now have the option to delegate parts of the query expression down to the DBMS and leave some to MyPL, e.g.~depending on runtime statistics or the likelihood that DB optimizations may even have a benefit. In addition, note that this implies that the \textbf{artificial boundary between an embedded MyPL-statement (a UDF) and the outer query is gone}, too.
From the user's point of view it is all just one big nested function. Which part of that function to execute where, e.g.~any nested expression, any outer wrapping expression, this becomes a holistic optimization task rather than two separate tasks where one is done by the PL-compiler in isolation and one by the DBMS in isolation. In fact, this decision is analogous to deciding which part of a program to run on a CPU (i.e.~in the PL) and which part to push to the GPU/TPU (i.e.~the DBMS). This opens up exciting future work on how to integrate these optimization and execution steps.

\noindent\textbf{(4.)~The QL is easily extensible.} Whether a function is defined by `a user' or by `a library', the QL should allow for using functions defined outside the realm of the database. The interesting question is \textit{where} such function will be executed. As discussed in (3.), a DBMS may decide to provide its own implementation of a function which was defined outside. One solution could be to think about function libraries that already give hints to the DBMS's query optimizer how to optimize them, i.e.~by providing rules and invariants. Like that we could avoid that each DBMS has to provide its own optimizations for those functions. 

\noindent\textbf{(5.)~The QL makes query injection impossible by definition.}
As we drop the idea of having a textual QL, there is no room for an injection attack like the infamous SQL injection~\cite{cwetop25}.
As the QL is purely functional, embedded in the PL, there is no way to change the boundary between the text belonging to the query and the text belonging to user-provided parameters.
If the user provides a parameter typed as a string containing functions trying to meddle with the outer function expression, that parameter will not be interpreted together with the `outer' query.
The parameter will be handled as a string not belonging to the query --- no matter what.
Thus, we get the same security guarantees as with existing database connectors splitting up query and parameters, e.g.~like in Psycopg, or preparing queries accordingly, i.e.~through prepared statements, or sanitizing user inputs for suspicious content (hint: don't do this, this can easily go wrong).
The big advantage of (3.) is, that even if the developer forgets to use the correct interfaces and/or other counter measures, in our approach, query injection simply cannot happen.

\noindent\textbf{(6.)~The QL never leaves the data model.} 
There are a couple of proposals that switch to other data models or serialization formats like arrays or JSON as workarounds for not being able to represent certain types of data~\cite{oracle,ong_sql++}. The same problem exists in ORMs, e.g.~in Django ORM, in order to phrase even a \texttt{GROUP}~ \texttt{BY}, the query does not return model instance(s) but dictionaries. A QL should not do this. The QL should strictly stay in the data model and distinguish between the data model (RMTM) and possible serialization formats (like JSON). 

\noindent\textbf{(7.)~The QL  guarantees snapshot semantics for the output.} To avoid data inconsistencies through post-query DB-traversals (see Section~\ref{sec:queryreturntype}), the QL should guarantee snapshot semantics. Similar to traditional snapshot semantics in an MVCC transaction, the QL must operates within the same consistent snapshot. Like that we provide a consistent view of the data. For any updates issued as part of that transaction, the standard MVCC validation mechanisms to provide serializability (not just snapshot isolation) apply.

\section{Map Views}
\label{sec:MapViews}

 In the following, we present our RMTM QL. It fulfills all seven requirements identified in~\autoref{sec:DBQL}. Our QL is based on map views which can be assembled into arbitrary expressions.

\begin{definition}[Map View.]
\label{def:DBView}
The general form of a Map View is:
$
V(\texttt{m}_{\text{in}})\mapsto \texttt{m}_{\text{out}}.
$
\end{definition}
In other words, the input of a map view is a map and the output is a map. So, technically, to be precise, we can write a view as a function signature using types as:
$
V(\texttt{RMT}_{\text{in}})\mapsto \texttt{RMT}_{\text{out}}.
$
However, for readability, in the following, we will stick to the term \textit{map views}. Just as traditional relational algebra operators operate on instances of relations (and not their types) our views operate on maps (the instance, not their type).
Note that in sharp contrast to relational algebra operators (e.g.~ $\pi$, $\sigma$, $\cup$, $\times$, $\bowtie$, ...), the input type \textbf{and} the return type of our map views \textbf{is a map of an RMT} which is not restricted to representing a single relation as in RM but can be any variant of a map (tuple, relation, database, set of databases, etc. recall~\autoref{fig:diffRMvsRMTM}).
Our approach also goes beyond the recently proposed \texttt{RESULTDB} extension~\cite{Nix2025ExtendingSQL} in that we cannot only return a relational subdatabase but can also return a set of databases plus a database with a completely different schema, e.g.~for aggregations.
Also note that for relational algebra there is often the confusion whether a relational algebra expression is supposed to be read as a declaration (not implying any order) or a function (implying a specific order).
By resorting to the term `views' we avoid this confusion from the beginning: we do not imply any specific order: we simply declare what map should be returned by a view.
For views defined on top of other views we also do not imply any order. You can interchangeably interpret `view' as (dynamic) `view'.

\subsection{Constraining Map Views}

\begin{definition}[Constraining Map View]
The general form of a Constraining Map View, i.e.~a Map View returning a subset of the maps and/or schema from the input $\texttt{m}_{\text{in}}$, is: $V_{\succ}(\texttt{m}_{\text{in}}, C)\mapsto \texttt{m}_{\text{out}}$
 where  $\text{schema}(\texttt{m}_{\text{out}}) \subseteq \text{schema}(\texttt{m}_{\text{in}})
\wedge  \texttt{m}_{\text{out}} \subseteq \texttt{m}_{\text{in}}$.
Here $C$ is a set of constraints defined on the input $\texttt{m}_{\text{in}}$.
\end{definition}

\begin{example}[Filter]~

\noindent\underline{\textbf{Query:}}
\noindent\textbf{Input Map:} RelMap Professors from Running Example

\noindent\textbf{Constraint:} Professors.age=42

\noindent\textbf{Output Map:} Professors reduced to qualifying maps

\noindent\textbf{Explanation:} simple filter on maps contained in Professors
\end{example}

\begin{example}[Project]~

\noindent\underline{\textbf{Query:}}
\noindent\textbf{Input Map:} Running Example DB Map

\noindent\textbf{Output Map:} give $\{$room, year$\}$

\noindent\textbf{Explanation:} all rooms used per year, duplicate-free.
\end{example}

\begin{example}[Yannakakis/Bernstein-DB Reduction (YBR)]~

\noindent\underline{\textbf{Query:}}
\noindent\textbf{Input Map:} Running Example DB Map

\noindent\textbf{Filter:} Professors.age=42, give.year=2025

\noindent\textbf{Output Map:} YBR-reduced Running Example DB Map

\noindent\textbf{Explanation:} no schema mapping/transformation: the output DB is a subset of the input DB (a DB reduction as defined by Yannakakis~\cite{yannakakis_acyclic-database-schemes}/Bernstein~\cite{bernstein_semi-joins}); RelMaps get reduced to contain only the (sub-)maps  contributing to the join.

\end{example}

\subsection{Transforming Map Views}

\begin{definition}[Transforming Map View]
The general form of a Transforming View, i.e.~a Map View returning transformed maps and/or schema from the input $\texttt{m}_{\text{in}}$, is: $V_{\looparrowright}(\texttt{m}_{\text{in}})\mapsto \texttt{m}_{\text{out}}$
where  $\text{schema}(\texttt{m}_{\text{out}}) \nsubseteq \text{schema}(\texttt{m}_{\text{in}})$.
\end{definition}

Transforming Map Views return a different output schema: that output is not simply a subset of the input: the input schema is transformed into something new.  This often implies that not only the schema but also the maps contained in the output map need to be transformed.  
\begin{example}[Grouping Sets]~

\noindent\underline{\textbf{Query:}}
\noindent\textbf{Input Map:} Running Example DB Map

\noindent\textbf{Output Map:} \{Agg1: $\{$give.year, count(*) $\}$, Agg2: $\{$give.room, count(*) $\}$\}

\noindent\textbf{Explanation:}  We output a map (a database) with two aggregation results: (1.)~a new map Agg1 showing the number of lectures per year, plus (2.)~a new map Agg2 showing the room usage per year.

In this example, we return multiple aggregates. This is similar to SQL's \texttt{GROUPING\;SETS}, however, in contrast to SQL, we do not need shoehorn the different aggregation results into a single relation with a common schema filling up missing attributes with \texttt{NULL}-values. 

\end{example}
Further use-cases of Transforming Map Views include: inner and outer joins (as they create a new schema for the joined tuples) but not semi-joins (as they keep the schema of one of the inputs).

\subsection{Using Map Views Out-of-place vs In-place}

Any map view can be used in two different ways:

\noindent\textbf{(1.)~Out-of-place:} This map view offers a different `perspective' on the input. It does not change the input map, i.e.~the underlying relation or database is not changed, but based on the input map ab out-of-place map view returns a snapshot in the sense of a copy of the data.  This corresponds to a classical read-only (\texttt{SELECT}) query copying data from the database to the outside but leaving the data in the database unchanged.

\noindent\textbf{(2.)~In-place:}  This map view describes a rule how to change the input map, i.e.~the underlying database. In this mode, the map view is used as a ``data rewrite rule''. Such a rule may mimic SQL's database state changing operations like \texttt{INSERT}/\texttt{UPDATE}/\texttt{DELETE}. However, in SQL, the latter operations are limited to tuples of a relation. In our approach, we can transform the underlying maps arbitrarily. With in-place map views the full expressive power of map views can be leveraged to transform the underlying data.

\subsection{SQL and RA are a Mix of Both}

Note, that any SQL-query $Q_{\text{SQL/RA}}$ as well as any corresponding relational algebra expression can be phrased as follows:
\begin{definition}[SQL/RA] The core structure of an SQL/RA-query is:
\label{def:SQL}
$
Q_{\text{SQL/RA}} :=  V_{\looparrowright,\succ}\big(\texttt{m}_{\text{in}}, C\big)\mapsto  \texttt{RV}_{\text{out}}.
$
\end{definition}

In other words, SQL and relational algebra perform \textit{both} of the constraining and transforming views as part of the same SQL/RA-statement. In addition, the output of a SQL/RA-query is restricted to an instance of RV, i.e.~a relation, and \textit{not} a DB.
 For instance, in an (inner or outer) join operation, SQL/RA blends both steps into one by identifying contributing tuples \textbf{and} transforming these tuples into new output tuples (and hence a new output schema).
 
 \newcounter{boxlblcounter}  
\newcommand{\makeboxlabel}[1]{(#1.)}
\newenvironment{boxlabel}
  {
  \begin{list}
    {\arabic{boxlblcounter}}
    {\usecounter{boxlblcounter}
     \setlength{\labelwidth}{1em}
     \setlength{\labelsep}{.3em}
     \setlength{\itemsep}{0pt}
     \setlength{\leftmargin}{0.9em}
     \setlength{\rightmargin}{0em}
     \setlength{\itemindent}{0em} 
     \let\makelabel=\makeboxlabel
    }
  }
{\end{list}
}


 \begin{table*}[!htbp]
  \renewcommand{\arraystretch}{1.17}
 \begin{footnotesize}
\begin{tabular}{| p{1.21cm} || p{1.4cm} |   p{3.3cm} | p{6.4cm} || p{3.5cm} |}
\hline
 \textbf{Map View Class}  &\textbf{Map View} \textbf{Subclass} &  \textbf{Explanation in RMTM Terminology} &\textbf{Use-Case in Traditional DB Terminology}  &  \textbf{SQL/RA Equivalent} \\\hline
\hline

	\multirow{3}{=}{\includegraphics[trim = 0mm 185mm 590mm 0mm, clip,width=.16\columnwidth,keepaspectratio,page=4]{Pics.pdf}}		
 &\textsf{partition} & \multirow{2}{=}{partition (aka group) an input map into multiple maps}&  split a relation into a DB of horizontal partitions (relations) & per partition views, physical~design\\\cline{4-5}
  &  &  &  split a DB into a set of fully co-partitioned DBs &  \textbf{not available}\\\cline{2-5}  	
  	& \textsf{replicate} & \multirow{2}{=}{replicate the input map into multiple maps}& replicate a relation into a DB of replica relations  &  \textbf{not available} \\\cline{4-5}
   &  &  &  replicate a DB into a DB of replica DBs & \textbf{not available} \\\cline{2-5} 
    	& \textsf{part\_rep} & \multirow{2}{=}{combination of partition and replicate, i.e.~partition with partial replication} & nested RAID-style partitioning and replication &  \textbf{not available} \\\cline{4-5}
             &  &  &  partition and replicate a DB into a set of fully co-partitioned and partially replicated DBs (e.g.~in a distributed DBMS) &  \textbf{not available} \\\cline{2-5}  
              &\multirow{3}{=}{\textsf{denormalize}, \textsf{factorize}} &  \multirow{2}{=}{factorize (denormalize) the input map} & split a relation into a normalized DB&  \textbf{not available}\\\cline{4-5}
             &  &  &  split a DB into a normalized DB &  \textbf{not available}\\\cline{4-5}  		
             &  &  &  split a join result into its factorized (co-partitioned) representation &  \textbf{not available}\\\cline{4-5}
\hline
\hline
	\multirow{3}{=}{\includegraphics[trim = 0mm 185mm 586mm 0mm, clip,width=.16\columnwidth,keepaspectratio,page=5]{Pics.pdf}}		
&  \textsf{cross\_product} & n-ary cross product creating concatenated output maps & compute the cross product on the input DB of relations and output concatenated maps& \multirow{2}{=}{\texttt{SELECT\;..\;FROM R, ..}} \\\cline{4-5}
             &  &  & compute the cross product on the input DB of DBs and output all pairs (e.g.~cross validation as in ML) &  \textbf{not available}\\\cline{2-5}
	& \textsf{join} & n-ary  join creating concatenated output maps & join the tuples of the relations in the input DB and output a single relation with the join results & \multirow{2}{=}{\texttt{SELECT\;..\;FROM R JOIN S ON ..}} \\\cline{4-5}
             &  &  & join the relations of the DBs in the input set of DBs and output a single DB with the join results (e.g.~as in Grace Hash Join) &   \textbf{not available}\\\cline{2-5}
 		& \textsf{semi\_join} & \multirow{2}{=}{n-ary semi join keeping the input schema} & compute the join on the input DB and return the subset of one of the inputs reduced to the maps participating in the join &\multirow{1}{=}{\texttt{SELECT\;..\;FROM R, ..}} \\\cline{2-5}
 		& \textsf{outer\_join} & \multirow{2}{=}{n-ary join keeping non-qualifying maps for some inputs} &  like \textsf{join} but also keep non-qualifying maps & \texttt{SELECT\;..\;FROM R OUTER JOIN S ON ..} \\\cline{4-5}		
       &  &  & dangling tuple identification for an entire input DB (not just a single relation) & \textbf{not available} \\\cline{2-5}
 		& op$_1$(op$_2$(  . ),  . ) & any other \textit{n-ary} relational algebra operator or expression & any traditional query & available, depends on expression\\\cline{2-5}
 		& \textsf{intersect} & \multirow{3}{=}{n-ary intersect of all inputs, requires input's schemas to be compatible} & intersect all relations in the input DB and return the result as a relation  &\texttt{INTERSECT} (only binary)\\\cline{4-5}
       &  &  & intersect all DBs in the input DB and return the result as a DB & \textbf{not available}\\\cline{2-5}
 		& \textsf{minus} &\multirow{2}{=}{n-ary minus of all inputs, requires input's schema(s) to be compatible} & remove all maps from first input relation that occur in second and following relations and return the result as a relation &\texttt{MINUS} (only binary)\\\cline{4-5}
		&  &  & remove all relations from first input DB that occur in second and following DB and return the result as a relation & \textbf{not available}\\\cline{2-5}
		& \textsf{union} & \multirow{2}{=}{integrate multiple inputs non-semantically}  & union all relations in the input DB and return the result as a relation &\texttt{UNION} (only binary)\\\cline{4-5}
		&  &  & union all DBs in the input DB and return the result as a relation & \textbf{not available}\\\cline{2-5}
\hline
\hline
	\multirow{3}{=}{\includegraphics[trim = 0mm 125mm 586mm 0mm, clip,width=.16\columnwidth,keepaspectratio,page=6]{Pics.pdf}}		
 & \textsf{project} & project the input &project a tuple & \texttt{SELECT\;..\;FROM R} \\\cline{4-4}
 &  &  & project a relation & \\\cline{4-5}
 &  &  & project a DB & \textbf{not available}\\\cline{4-5}
 &  &  & project a set of DBs & \textbf{not available} \\\cline{2-5}
					& \textsf{compute} & based on the input map, compute an output map  & variants of \textsf{project}& similar to  \textsf{project}\\\cline{2-5}
 		& \textsf{rename} &\multirow{2}{=}{rename the key in a key/value-mapping of an RT} & in a relation, rename the row-id of tuple & \textbf{not available}\\\cline{4-5}
 &  &  & in a DB, rename the relation-id of a relation & \texttt{ALTER TABLE .
RENAME TO}\\\cline{4-5}
 &  &  & in a set of DBs, rename the database-id of a DB &\texttt{ALTER DATABASE .
RENAME TO} \\\cline{2-5}
 		& \textsf{aggregate} & \multirow{1}{=}{combine multiple/all input maps} & aggregate tuples in input relation to tuples in output relation & \texttt{SELECT agg() FROM R} \\\cline{4-5}
 &  &  & aggregate relations in input DB to relation(s) in output DB & \textbf{not available}\\\cline{4-5}
 &  &  & aggregate DBs in input set of DB to DBs in output DB &  \textbf{not available}\\\cline{2-5}
 		& op$_1$(op$_2$( . )) & any other \textit{unary} relational algebra operator or expression & any traditional relational algebra operator or expression with a single input relation& available, depends on expression\\\cline{2-5}
		& \textsf{insert} & \multirow{1}{=}{add maps; inverse to \textsf{filter}/\textsf{delete}} & add tuples to relation 	&\texttt{INSERT INTO} \\\cline{4-5}
		&  &  & add relations to DB	& \texttt{CREATE TABLE}, only one table \\\cline{4-5}
		&  &  & add DBs to set of DBs 	&  \texttt{CREATE DATABASE}\\\cline{2-5}
 		& \textsf{update} & change maps & update the tuples of a relation	&\texttt{UPDATE} \\\cline{4-5}
		&  &  & update the relations of a DB	&\texttt{ALTER TABLE}, only one table \\\cline{4-5}
		&  &  & update the DBs of a set of DBs	&\textbf{not available} \\\cline{2-5}
	& \textsf{filter}/\textsf{delete}  &  \multirow{1}{=}{remove maps from the input} &delete/filter tuples from relation	&\texttt{DELETE}, \texttt{WHERE}, \texttt{HAVING} \\\cline{4-5}
		&  &  & drop relations from DB	&\texttt{DROP\;TABLE}\\\cline{4-5}
		&  &  & delete DBs from set of DBs	&\texttt{DROP\;DATABASE} \\\cline{2-5}
	&  \textsf{subDB\_inner}& return a subDB   & execute Yannakakis~\cite{yannakakis_acyclic-database-schemes}/Bernstein~\cite{bernstein_semi-joins}-reduction (YBR)&\multirow{3}{=}{\textbf{not available}, but extension of~\cite{Nix2025ExtendingSQL}: \texttt{SELECT}~\texttt{RESULTDB ..}}\\\cline{4-4}
	&  &  & YBR in-place: remove dangling tuples from input database&\\\cline{2-4}
	& \textsf{subDB\_outer} &  return a subDB with outer& like \textsf{subDB\_inner}, but for parts of the input return outer maps & \\\cline{2-4}
\hline
\end{tabular}
\caption{A Foundational Taxonomy of Map Views\label{tab:principalmapViews}. Due to space constraints this table focusses on use-cases not existing in RA/SQL.
Keep in mind that all Map Views operate on an arbitrary input map and output an arbitrary output map. Hence the input and output can be any combination of a tuple, a relation, a DB or a set of DBs, ... (i.e., this is at least $4^2=16$ variants)
}
\end{footnotesize}
\vspace*{-1.5em}
\end{table*}
  \renewcommand{\arraystretch}{1}

\section{Three Classes of Map Views, Their Subclasses, and their Use-Cases}
\label{sec:threeclassesofEVs}

Given an input map and an output map, what are the principal types of map views that exist?

\begin{definition}[Increase View]
An Increase View transforms an input map $\texttt{m}_{\text{in}}$ into an output map $\texttt{m}_{\text{out}}$:
$
\textsf{increase}( \texttt{m}_{\text{in}}) \mapsto \texttt{m}_{\text{out}}
$
such that $\textsf{order}(\texttt{m}_{\text{in}}) < \textsf{order}(\texttt{m}_{\text{out}})$.
\label{def:increaseview}
In other words, the output map has a higher order (see~\autoref{def:order}) than the input map.
\end{definition}
\begin{definition}[Decrease View]
A Decrease View transforms an input map $\texttt{m}_{\text{in}}$ into an output map $\texttt{m}_{\text{out}}$:
$
\textsf{decrease}(\texttt{m}_{\text{in}}) \mapsto  \texttt{m}_{\text{out}}
$
such that $\textsf{order}(\texttt{m}_{\text{in}}) > \textsf{order}(\texttt{m}_{\text{out}})$.
\label{def:decreaseview}
In other words, the output map has a lower order than the input map.
\end{definition}

An increase view may have an inverse decrease view
$
\textsf{decrease}\big(\textsf{increase}( \texttt{m}_{\text{in}})\big) \mapsto \texttt{m}_{\text{in}}
$
and vice versa. 

\begin{definition}[Map View]
A map view transforms an input map $\texttt{m}_{\text{in}}$ into an output map $\texttt{m}_{\text{out}}$ :
$
\textsf{map}(\texttt{m}_{\text{in}}) \mapsto \texttt{m}_{\text{out}}
$
such that $\textsf{order}(\texttt{m}_{\text{in}}) = \textsf{order}(\texttt{m}_{\text{out}})$.
\label{def:mapview}
In other words, the output map has the same order as the input map.
\end{definition}
Table~\ref{tab:principalmapViews}
lists important instances of these three classes of Map Views as well as prominent use-cases. Due to space constraints we can only show a subset of what is possible with map views. In the table, we therefore focus on use-cases going beyond RA/SQL.

For each of the three classes of map views, we show important map view subclasses. For each of these subclasses, we explain that subclass using our RMTM terminology, and then show the various applications of that subclass using traditional database terminology. We also list the SQL/RA counterparts for the cases where they exist. 

\section{Related Work}
\label{sec:RW}

\noindent\textbf{Most Relevant Related Work.} See our discussion of~\cite{Deshpande} and ~\cite{Nix2025ExtendingSQL} in the Introduction.

\noindent\textbf{Obviously Related Work.} In terms of relational QLs there were other proposals that basically boil down to proposing a hybrid of RA operators and SQL-style syntax in the PL~\cite{piglatin,linq,PRQL,ShuteBBBDKLMMSWWY24}; a variant of RA  in the PL, e.g.~XXL~\cite{BBD+01}; or offer both RA and SQL~\cite{ZahariaCDDMMFSS12}, and/or combine that with a pipe syntax~\cite{PRQL,ShuteBBBDKLMMSWWY24}. The most recent call to replace SQL with something more functional is~\cite{NL24}. Again, that work builds on RA and allows developers to build pipelines through a dot syntax, very similar to Django ORM QuerySets. Such a syntax does work on RM limited to binary operators outputting one  relation. However, for RMTM, and its ability to input and output entire databases a dot syntax simply does not work. Moreover, in that work, the underlying RM and its limitations are not questioned.
There were a couple of other approaches trying to break the RA operator abstraction into suboperators~\cite{Lohman88,DittrichN20,BandleG21,KohnL021}. 
In fact, any of our map views doing `less' than producing a single output relation can be coined a \textbf{suboperator}. Similarly, any map view returning something bigger than a relation could be called a \textbf{superoperator}.
Another line of work tries to marry the relational world with JSON, e.g.~SQL++\cite{ong_sql++} and Oracle~\cite{oracle}. However, as outlined by~\cite{Nix2025ExtendingSQL}, these approaches cannot represent N:M result sets. All these approaches mix up the underlying conceptual data model with its representation.
Rel~\cite{rel} has a very similar motivation as our work. However, it keeps the abstraction of relations to be sets of tuples and tuples to be ``an ordered immutable sequence of data values''~\cite{rel} which we both discard to be both key-value mappings. In addition, from the description on the Rel website it is unclear how attributes are typed. Syntax-wise, Rel is quite different from the most common PLs like Rust, C++, Python, etc. and thus is much harder to integrate into those PLs. 

\noindent\textbf{Somewhat Related Work.}  Since the dawn of the relational model~\cite{codd_relational-model}, SQL~\cite{ChamberlinB74}, and the entity Relationship-model~\cite{Chen75}, there has been criticism, e.g.~\cite{date_critique-sql,codd_normalization,NL24,ShuteBBBDKLMMSWWY24}.
There were a couple of unification approaches, however, yet we are not aware of any proposal that brings both the relational and object-oriented world as much in sync and truely unifies them as in our paper. In the 80ies, a number of interesting proposals were made to develop query languages directly on ERM~\cite{ParentS84,ElmasriW81,CampbellEC85,CzejdoERE90}. Yet all of these works fall behind our approach in terms of expressiveness.

\section{Experiments}
\label{sec:experiments}

\label{sec:exp:swizzlingvsjoins}

In this section we show experimental results of the two star join algorithms discussed in~\autoref{sec:qprocandopt}.  The point of this experiment is to show the  impact of keeping pointers vs foreign keys in RM.

\newcommand\myscale{0.518}
\begin{figure*}[t!]
  \centering
  \begin{subfigure}[t]{\myscale\columnwidth}
    \centering
    \includegraphics[scale=0.25]{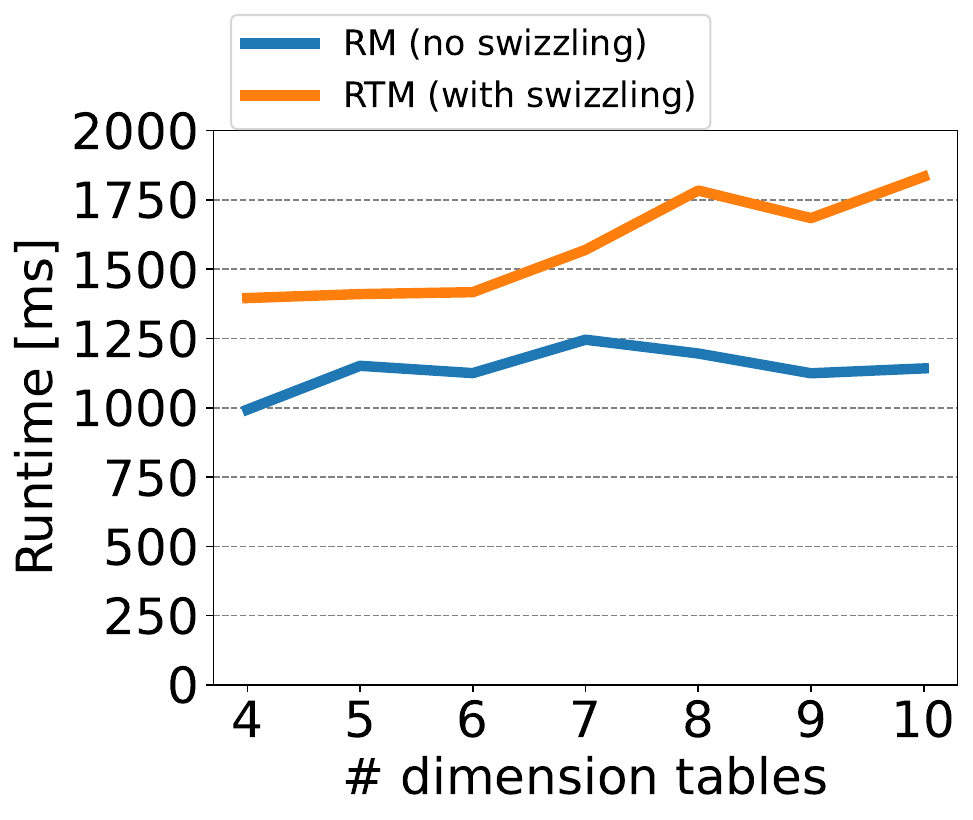}
    \caption{Fact table build times}
    \label{fig:star-swizzling-build-times}
  \end{subfigure}
  \begin{subfigure}[t]{\myscale\columnwidth}
    \centering
    \includegraphics[scale=0.25]{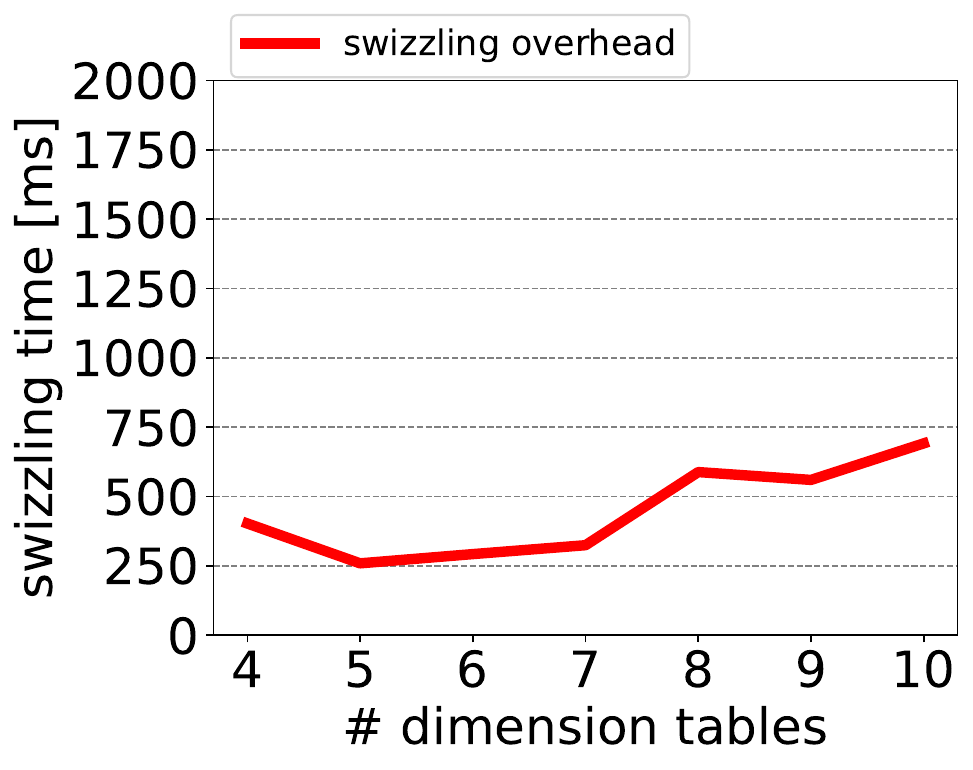}
    \caption{Swizzling overheads of RMTM for initial load (may be overlapped with I/O)}
    \label{fig:star-swizzling-overheads}
  \end{subfigure}
  \begin{subfigure}[t]{\myscale\columnwidth}
    \centering
    \includegraphics[scale=0.25]{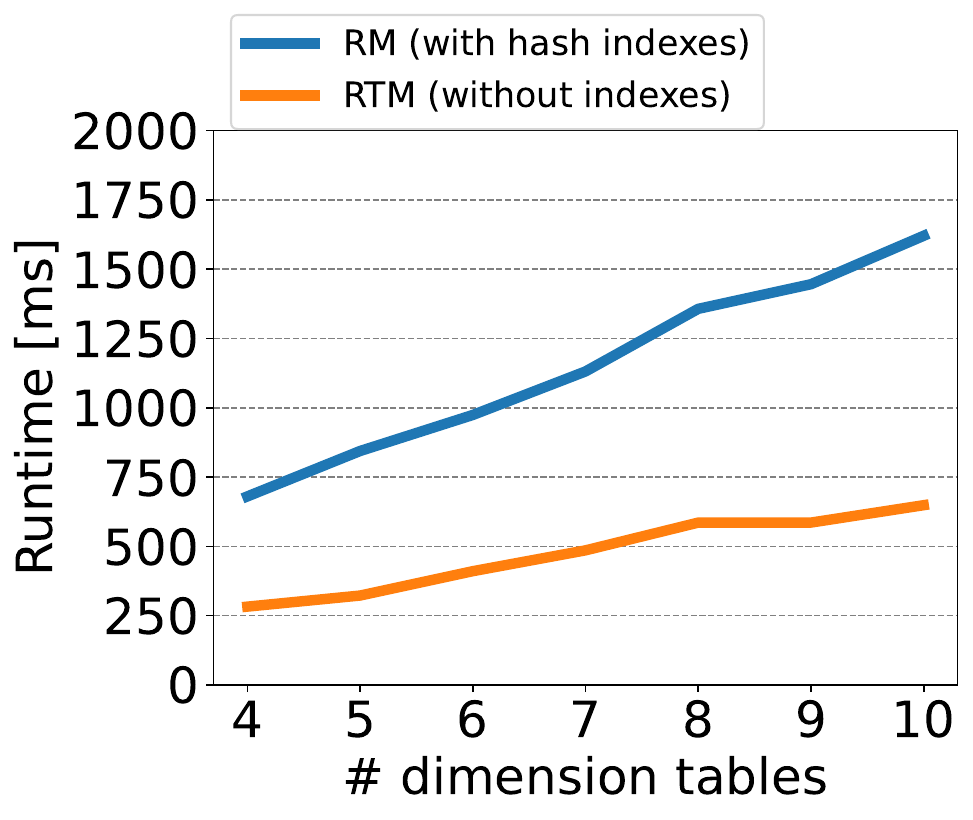}
    \caption{Star join runtimes}
    \label{fig:star-runtimes}
  \end{subfigure}
  \begin{subfigure}[t]{\myscale\columnwidth}
    \centering
    \includegraphics[scale=0.25]{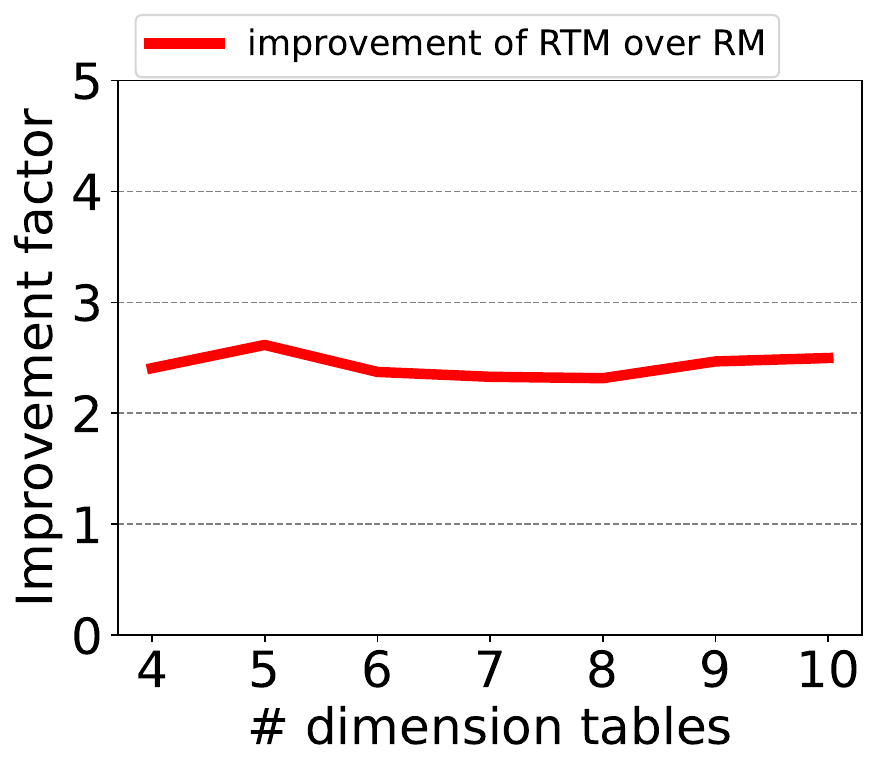}
    \caption{Improvement factor for star join runtime}
    \label{fig:star-improvements}
  \end{subfigure}

  \caption{Results of a star join on a fact table with 50~million tuples varying the number of dimension tables: RM vs RMTM}
  \label{fig:starjoin}

\end{figure*}

\noindent\textbf{Setup.} Both join algorithms were implemented in C++  and compiled with \texttt{clang++\;-O2}. All experiments were run on an M3 Max with 48GB main memory, single-threaded. All data resided in main memory. All experiments were repeated 10 times and we report the median of these repetitions. We created a star schema with 50M tuples in the fact table and 100 tuples in each dimension table. All data is uniformly distributed. All code including the notebooks to create the plots is available at \url{https://anonymous.4open.science/r/4242/}.

\noindent\textbf{Results.} \autoref{fig:star-runtimes} shows the runtime of both algorithms when increasing the number of dimension tables. For both algorithms, we observe a linear behavior. This is expected as the effort for joining tuples from the fact table is linear in the number of dimension tables.  However, the striking difference is the constants in these algorithms: in RM, we have to use the hash indexes to concatenate/reconstruct tuples. In contrast, in RMTM, we can directly look up tuples in the dimension tables by following the pointers present in tuples in the fact table. Like that we save a few extra hops for each lookup. \autoref{fig:star-improvements} shows the performance improvement of RMTM over RM: it is approximately a factor~3.

As pointed out in~\autoref{sec:qprocandopt}, the price we pay is swizzling costs when loading the data. However, that price only has to be paid once when loading the data from SSD/disk\footnote{Another option here is to swizzle the data not when loading but only when the first query touches it.}. 
\autoref{fig:star-swizzling-build-times} explores these costs.
We see the result for uniformly randomly instantiating 50M tuples in the fact table based on the existing dimension tables.
For RM, this is simply a nested loop iterating over the dense integer keys used for the dimension tables: This is a lower bound for these costs (which would in reality involve looking up dimension tables and their keys).
For RMTM, we do the same thing but additionally create pointers to the concrete tuples in the dimension tables: this is obviously more costly.
Yet, the difference of the two algorithms represents very well the costs for swizzling, i.e.~the costs of flipping foreign keys to pointers. \autoref{fig:star-swizzling-overheads} shows the results:
For RMTM, we observe overheads in the range of 250--700ms for 10~dimension tables. 
Yet, even for a single query using RMTM, we save 1000ms of work (\autoref{fig:star-runtimes}).
Therefore, the little additional overhead pays off dramatically at query time.
In fact, the pointers behave in a way like a perfect index.

These initial (and we believe very motivating) results obviously call for a more detailed E\&A-study.
However, due to the abundance of material in this paper and the space constraints of 12~pages, we cannot go deeper here and leave it to future work.

\section{Conclusions: Getting Rid of...}
\label{sec:gettingrid}

We hope that this paper triggers an honest and open discussion in the database community and beyond about our $\sim$55 year old relational history. SQL, RA, RM, ERM, and ORMs have conquered the world.  But it is time for a checkpoint. 

\subsection{SQL and Relational Algebra?}

SQL produces a flat, tabular view on the data. Each \texttt{SELECT}-query produces another data island. In addition, in SQL data filtering and transformation are heavily intertwined (see Definition~\ref{def:SQL}).
This is legacy.
And it creates a lot of artificial problems.
We believe that in 2025, SQL is not adequate anymore.
Therefore, a functional language consisting of Map Views inputting and outputting maps of arbitrary granularity is much more appropriate, and much more expressive and powerful.
Map Views are much more powerful than RA and can represent RA as a small special case. And SQL injection~\cite{cwetop25} becomes impossible --- for free. 

Strictly speaking, we believe that we should get rid of the idea of having a query language disconnected from the PL.
We need PL counterparts representing our Map Views.
Whatever their syntax is in a PL, the job description of the DBMS is to take these expressions and suggest efficient query plans and execute them.
In terms of software engineering, our proposal calls for an interface where the different Map Views are implemented by each PL: as stubs/placeholders \textit{and} with DB implementations/optimizations.
Thus, these views can always be executed with and without the DBMS.

In the long run, SQL should evolve into becoming the second best way of querying and manipulating data, just like COBOL became outdated but is still around here and there. We should drop the notion in our database evangelists' heads that SQL or SQL-like or any other textual languages must remain the principal query language -- just because SQL dominates the world today for historic reasons.

In summary, \textbf{we should get rid of SQL and RA and replace them by map views}. 
\subsection{The Relational Model?}

The problem with RM are its historical constraints and the misunderstandings that come with that. RMTM is like RM on (a lot of) steroids, and without all the legacy used by Codd~\cite{codd_relational-model}.
Plus, we do not define `a relation' to be a subset of some cartesian product of domains, etc.
And, we do not model a `tuple' as `a sequence of attribute values'. Instead, we use maps for all granularities.
A collection of these maps is then put into a higher order map which enforces certain constraints on these maps, e.g.~by enforcing the existence of certain keys.
Like that we can represent the traditional tuples, relations, databases, and set of databases with a single abstraction on all those levels: a map. Plus we constrain these maps arbitrarily with Map Types.

In summary, \textbf{we should get rid of RM and replace it by RMTM}.

\subsection{Entity-Relationship Model?}

ERM, in particular as a visual data modeling tool, can be really helpful when getting started with data modeling. 
We should not be religious about ERM's limitations and whenever possible allow for using the more flexible RMTM. The right, practical trade-off among the two, and what that means in terms of a visual modeling tool needs to be determined as part of future work.

In summary,  \textbf{ERM is a useful visual data modeling tool}.

\subsection{ORMs?}
\label{sec:gettingrid:ORMs}

Our functional QL can be used directly rather than using workarounds like e.g.~Django ORM which is still mapped to SQL and thus produces a flat representation of the data. With our QL there is no impedance mismatch.
Thus, the need for using an ORM in the first place, goes away. What is left is a super strong feature of ORMs, and a very fragile security flaw of SQL: as SQL is text-based and then parsed, SQL is easily attackable by \textit{SQL injection}.
Shockingly, in 2024, SQL injection was still the third most dangerous software weakness~\cite{cwetop25}.
By switching to a functional (non-textual) language this problem goes away --- for free. So, even this last reason to use an ORM is gone.

There are even more benefits of this: when using an ORM we keep two different database schemas: one relational database schema inside the RDBMS, and one object-oriented schema in the ORM, e.g.~in Django ORM this schema is declared in a \texttt{models.py}-file.
Both schemas have to be kept in sync through so-called \textit{migrations}, i.e.~when we change the \texttt{models.py}-file, we have to manually trigger a schema sync-operation by calling \texttt{makemigrations} and \texttt{migrate} which will then change the RDBMS schema accordingly.
And you must not change the RDBMS schema yourself as that will confuse the ORM\footnote{Django ORM offers options to allow developers to change, on a per table-basis, either the schema in the RDBMS, in the ORM, or in both. Still, it is two places that need to be synced to make the ORM work correctly.}.
Depending on the features offered by the DBMS, migrations can go wrong\footnote{For instance, as of writing this, Django ORM still does not support composite primary keys.
Django was first released in 2005.
Twenty years later, as of writing these lines, the stable version of Django is 5.1.7. Yet \href{https://docs.djangoproject.com/en/dev/tocomposite-primary-key/}{composite primary} keys are only planned for~5.2.}.
Therefore, you must make sure that whatever schema change you did in the ORM, these changes can then also be executed in your production RDBMS (or vice versa)\footnote{It is a good idea to verify this automatically as part of your CI.}.

With our approach, it is enough to keep the schema in one place. Hence, \textbf{we only need one schema declaration and thus, we do not need these migrations}. As a consequence, one source of errors is gone.

In summary, \textbf{we can get rid of ORMs}.

\vspace*{1em}

\noindent\textbf{Acknowledgments.} We did \textbf{not} use ChatGPT to generate text or ideas for this paper. However we used ChatGPT to help create tikz-code, fix math notation, and help refactor and comment the C++ code for the experiments. 

\bibliographystyle{ACM-Reference-Format}
\bibliography{main}


\begin{thebibliography}{35}


\ifx \showCODEN    \undefined \def \showCODEN     #1{\unskip}     \fi
\ifx \showDOI      \undefined \def \showDOI       #1{#1}\fi
\ifx \showISBNx    \undefined \def \showISBNx     #1{\unskip}     \fi
\ifx \showISBNxiii \undefined \def \showISBNxiii  #1{\unskip}     \fi
\ifx \showISSN     \undefined \def \showISSN      #1{\unskip}     \fi
\ifx \showLCCN     \undefined \def \showLCCN      #1{\unskip}     \fi
\ifx \shownote     \undefined \def \shownote      #1{#1}          \fi
\ifx \showarticletitle \undefined \def \showarticletitle #1{#1}   \fi
\ifx \showURL      \undefined \def \showURL       {\relax}        \fi
\providecommand\bibfield[2]{#2}
\providecommand\bibinfo[2]{#2}
\providecommand\natexlab[1]{#1}
\providecommand\showeprint[2][]{arXiv:#2}

\bibitem[auto prefetch(2025)]%
        {djangoautoprefetch}
\bibfield{author}{\bibinfo{person}{Django auto prefetch}.}
  \bibinfo{year}{2025}\natexlab{}.
\newblock \bibinfo{title}{{Django auto prefetch}}.
\newblock
  \bibinfo{howpublished}{\url{https://pypi.org/project/django-auto-prefetch}}.
\newblock
\newblock
\shownote{[Online; accessed 13-March-2025]}.


\bibitem[Bandle and Giceva(2021)]%
        {BandleG21}
\bibfield{author}{\bibinfo{person}{Maximilian Bandle} {and}
  \bibinfo{person}{Jana Giceva}.} \bibinfo{year}{2021}\natexlab{}.
\newblock \showarticletitle{Database Technology for the Masses: Sub-Operators
  as First-Class Entities}.
\newblock \bibinfo{journal}{\emph{Proc. {VLDB} Endow.}} \bibinfo{volume}{14},
  \bibinfo{number}{11} (\bibinfo{year}{2021}), \bibinfo{pages}{2483--2490}.
\newblock


\bibitem[Bernstein and Chiu(1981)]%
        {bernstein_semi-joins}
\bibfield{author}{\bibinfo{person}{Philip~A. Bernstein} {and}
  \bibinfo{person}{Dah{-}Ming~W. Chiu}.} \bibinfo{year}{1981}\natexlab{}.
\newblock \showarticletitle{Using Semi-Joins to Solve Relational Queries}.
\newblock \bibinfo{journal}{\emph{J. {ACM}}} \bibinfo{volume}{28},
  \bibinfo{number}{1} (\bibinfo{year}{1981}), \bibinfo{pages}{25--40}.
\newblock
\urldef\tempurl%
\url{https://doi.org/10.1145/322234.322238}
\showDOI{\tempurl}


\bibitem[Campbell et~al\mbox{.}(1985)]%
        {CampbellEC85}
\bibfield{author}{\bibinfo{person}{Douglas~M. Campbell},
  \bibinfo{person}{David~W. Embley}, {and} \bibinfo{person}{Bogdan~D. Czejdo}.}
  \bibinfo{year}{1985}\natexlab{}.
\newblock \showarticletitle{A Relationally Complete Query Language for an
  Entity-Relationship Model}. In \bibinfo{booktitle}{\emph{Entity-Relationship
  Approach: The Use of {ER} Concept in Knowledge Representation, Proceedings of
  the Fourth International Conference on Entity-Relationship Approach, Chicago,
  Illinois, USA, 29-30 October 1985}},
  \bibfield{editor}{\bibinfo{person}{Peter~P. Chen}} (Ed.).
  \bibinfo{publisher}{{IEEE} Computer Society and North-Holland},
  \bibinfo{pages}{90--97}.
\newblock


\bibitem[Chamberlin and Boyce(1974)]%
        {ChamberlinB74}
\bibfield{author}{\bibinfo{person}{Donald~D. Chamberlin} {and}
  \bibinfo{person}{Raymond~F. Boyce}.} \bibinfo{year}{1974}\natexlab{}.
\newblock \showarticletitle{{SEQUEL:} {A} Structured English Query Language}.
  In \bibinfo{booktitle}{\emph{Proceedings of 1974 {ACM-SIGMOD} Workshop on
  Data Description, Access and Control, Ann Arbor, Michigan, USA, May 1-3,
  1974, 2 Volumes}}, \bibfield{editor}{\bibinfo{person}{Gene Altshuler},
  \bibinfo{person}{Randall Rustin}, {and} \bibinfo{person}{Bernard~D. Plagman}}
  (Eds.). \bibinfo{publisher}{{ACM}}, \bibinfo{pages}{249--264}.
\newblock


\bibitem[Chen(1975)]%
        {Chen75}
\bibfield{author}{\bibinfo{person}{Peter~P. Chen}.}
  \bibinfo{year}{1975}\natexlab{}.
\newblock \showarticletitle{The Entity-Relationship Model: Toward a Unified
  View of Data}. In \bibinfo{booktitle}{\emph{Proceedings of the International
  Conference on Very Large Data Bases, September 22-24, 1975, Framingham,
  Massachusetts, {USA}}}, \bibfield{editor}{\bibinfo{person}{Douglas~S. Kerr}}
  (Ed.). \bibinfo{publisher}{{ACM}}, \bibinfo{pages}{173}.
\newblock


\bibitem[Codd(1970)]%
        {codd_relational-model}
\bibfield{author}{\bibinfo{person}{E.~F. Codd}.}
  \bibinfo{year}{1970}\natexlab{}.
\newblock \showarticletitle{A Relational Model of Data for Large Shared Data
  Banks}.
\newblock \bibinfo{journal}{\emph{Commun. {ACM}}} \bibinfo{volume}{13},
  \bibinfo{number}{6} (\bibinfo{year}{1970}), \bibinfo{pages}{377--387}.
\newblock
\urldef\tempurl%
\url{https://doi.org/10.1145/362384.362685}
\showDOI{\tempurl}


\bibitem[Codd(1971)]%
        {codd_normalization}
\bibfield{author}{\bibinfo{person}{E.~F. Codd}.}
  \bibinfo{year}{1971}\natexlab{}.
\newblock \showarticletitle{Further Normalization of the Data Base Relational
  Model}.
\newblock \bibinfo{journal}{\emph{Research Report / {RJ} / {IBM} / San Jose,
  California}}  \bibinfo{volume}{{RJ909}} (\bibinfo{year}{1971}).
\newblock


\bibitem[Czejdo et~al\mbox{.}(1990)]%
        {CzejdoERE90}
\bibfield{author}{\bibinfo{person}{Bogdan~D. Czejdo}, \bibinfo{person}{Ramez
  Elmasri}, \bibinfo{person}{Marek Rusinkiewicz}, {and}
  \bibinfo{person}{David~W. Embley}.} \bibinfo{year}{1990}\natexlab{}.
\newblock \showarticletitle{A Graphical Data Manipulation Language for an
  Extended Entity-Relationship Model}.
\newblock \bibinfo{journal}{\emph{Computer}} \bibinfo{volume}{23},
  \bibinfo{number}{3} (\bibinfo{year}{1990}), \bibinfo{pages}{26--36}.
\newblock
\urldef\tempurl%
\url{https://doi.org/10.1109/2.50270}
\showURL{%
\tempurl}


\bibitem[Date(1984)]%
        {date_critique-sql}
\bibfield{author}{\bibinfo{person}{C.~J. Date}.}
  \bibinfo{year}{1984}\natexlab{}.
\newblock \showarticletitle{A Critique of the {SQL} Database Language}.
\newblock \bibinfo{journal}{\emph{{SIGMOD} Rec.}} \bibinfo{volume}{14},
  \bibinfo{number}{3} (\bibinfo{year}{1984}), \bibinfo{pages}{8--54}.
\newblock
\urldef\tempurl%
\url{https://doi.org/10.1145/984549.984551}
\showDOI{\tempurl}


\bibitem[Date(2000)]%
        {date}
\bibfield{author}{\bibinfo{person}{C.~J. Date}.}
  \bibinfo{year}{2000}\natexlab{}.
\newblock \bibinfo{booktitle}{\emph{An introduction to database systems {(7.}
  ed.)}}.
\newblock \bibinfo{publisher}{Addison-Wesley-Longman}.
\newblock


\bibitem[Date et~al\mbox{.}(2002)]%
        {DDL02}
\bibfield{author}{\bibinfo{person}{C.~J. Date}, \bibinfo{person}{Hugh Darwen},
  {and} \bibinfo{person}{Nikos~A. Lorentzos}.} \bibinfo{year}{2002}\natexlab{}.
\newblock \bibinfo{booktitle}{\emph{Temporal data and the relational model}}.
\newblock \bibinfo{publisher}{Elsevier}.
\newblock


\bibitem[Dehspande(2025)]%
        {Deshpande}
\bibfield{author}{\bibinfo{person}{Amol Dehspande}.}
  \bibinfo{year}{2025}\natexlab{}.
\newblock \showarticletitle{Beyond Relations: A Case for Elevating to the
  Entity-Relationship Abstraction}.
\newblock \bibinfo{journal}{\emph{{CIDR} '25}} (\bibinfo{year}{2025}).
\newblock


\bibitem[den Bercken et~al\mbox{.}(2001)]%
        {BBD+01}
\bibfield{author}{\bibinfo{person}{Jochen~Van den Bercken},
  \bibinfo{person}{Bj{\"o}rn Blohsfeld}, \bibinfo{person}{Jens-Peter Dittrich},
  \bibinfo{person}{J{\"u}rgen Kr{\"a}mer}, \bibinfo{person}{Tobias
  Sch{\"a}fer}, \bibinfo{person}{Martin Schneider}, {and}
  \bibinfo{person}{Bernhard Seeger}.} \bibinfo{year}{2001}\natexlab{}.
\newblock \showarticletitle{{XXL - A Library Approach to Supporting Efficient
  Implementations of Advanced Database Queries}}. In
  \bibinfo{booktitle}{\emph{VLDB}}.
\newblock


\bibitem[Dittrich and Nix(2020)]%
        {DittrichN20}
\bibfield{author}{\bibinfo{person}{Jens Dittrich} {and} \bibinfo{person}{Joris
  Nix}.} \bibinfo{year}{2020}\natexlab{}.
\newblock \showarticletitle{The Case for Deep Query Optimisation}. In
  \bibinfo{booktitle}{\emph{10th Conference on Innovative Data Systems
  Research, {CIDR} 2020, Amsterdam, The Netherlands, January 12-15, 2020,
  Online Proceedings}}. \bibinfo{publisher}{www.cidrdb.org}.
\newblock


\bibitem[Elmasri and Wiederhold(1981)]%
        {ElmasriW81}
\bibfield{author}{\bibinfo{person}{Ramez Elmasri} {and} \bibinfo{person}{Gio
  Wiederhold}.} \bibinfo{year}{1981}\natexlab{}.
\newblock \showarticletitle{{GORDAS:} {A} Formal High-Level Query Language for
  the Entity-Relationship Model}. In
  \bibinfo{booktitle}{\emph{Entity-Relationship Approach to Information
  Modeling and Analysis, Proceedings of the Second International Conference on
  the Entity-Relationship Approach (ER'81), Washington, DC, USA, October 12-14,
  1981}}, \bibfield{editor}{\bibinfo{person}{Peter~P. Chen}} (Ed.).
  \bibinfo{publisher}{North-Holland}, \bibinfo{pages}{49--72}.
\newblock


\bibitem[Grant(2008)]%
        {grant_null-values}
\bibfield{author}{\bibinfo{person}{John Grant}.}
  \bibinfo{year}{2008}\natexlab{}.
\newblock \showarticletitle{Null values in {SQL}}.
\newblock \bibinfo{journal}{\emph{{SIGMOD} Rec.}} \bibinfo{volume}{37},
  \bibinfo{number}{3} (\bibinfo{year}{2008}), \bibinfo{pages}{23--25}.
\newblock
\urldef\tempurl%
\url{https://doi.org/10.1145/1462571.1462575}
\showDOI{\tempurl}


\bibitem[Kemper and Kossmann(1993)]%
        {KemperK93}
\bibfield{author}{\bibinfo{person}{Alfons Kemper} {and} \bibinfo{person}{Donald
  Kossmann}.} \bibinfo{year}{1993}\natexlab{}.
\newblock \showarticletitle{Adaptable Pointer Swizzling Strategies in Object
  Bases}. In \bibinfo{booktitle}{\emph{Proceedings of the Ninth International
  Conference on Data Engineering, April 19-23, 1993, Vienna, Austria}}.
  \bibinfo{publisher}{{IEEE} Computer Society}, \bibinfo{pages}{155--162}.
\newblock


\bibitem[Kohn et~al\mbox{.}(2021)]%
        {KohnL021}
\bibfield{author}{\bibinfo{person}{Andr{\'{e}} Kohn}, \bibinfo{person}{Viktor
  Leis}, {and} \bibinfo{person}{Thomas Neumann}.}
  \bibinfo{year}{2021}\natexlab{}.
\newblock \showarticletitle{Building Advanced {SQL} Analytics From Low-Level
  Plan Operators}. In \bibinfo{booktitle}{\emph{{SIGMOD} '21: International
  Conference on Management of Data, Virtual Event, China, June 20-25, 2021}},
  \bibfield{editor}{\bibinfo{person}{Guoliang Li}, \bibinfo{person}{Zhanhuai
  Li}, \bibinfo{person}{Stratos Idreos}, {and} \bibinfo{person}{Divesh
  Srivastava}} (Eds.). \bibinfo{publisher}{{ACM}}, \bibinfo{pages}{1001--1013}.
\newblock


\bibitem[Linq(2025)]%
        {linq}
\bibfield{author}{\bibinfo{person}{Linq}.} \bibinfo{year}{2025}\natexlab{}.
\newblock \bibinfo{title}{{Linq}}.
\newblock
  \bibinfo{howpublished}{\url{https://learn.microsoft.com/en-us/dotnet/csharp/linq/}}.
\newblock
\newblock
\shownote{[Online; accessed 1-April-2025]}.


\bibitem[Lohman(1988)]%
        {Lohman88}
\bibfield{author}{\bibinfo{person}{Guy~M. Lohman}.}
  \bibinfo{year}{1988}\natexlab{}.
\newblock \showarticletitle{Grammar-like Functional Rules for Representing
  Query Optimization Alternatives}. In \bibinfo{booktitle}{\emph{Proceedings of
  the 1988 {ACM} {SIGMOD} International Conference on Management of Data,
  Chicago, Illinois, USA, June 1-3, 1988}},
  \bibfield{editor}{\bibinfo{person}{Haran Boral} {and}
  \bibinfo{person}{Per{-}{\AA}ke Larson}} (Eds.). \bibinfo{publisher}{{ACM}
  Press}, \bibinfo{pages}{18--27}.
\newblock


\bibitem[Mitre(2024)]%
        {cwetop25}
\bibfield{author}{\bibinfo{person}{Mitre}.} \bibinfo{year}{2024}\natexlab{}.
\newblock \bibinfo{title}{{2024 CWE Top 25 Most Dangerous Software
  Weaknesses}}.
\newblock
  \bibinfo{howpublished}{\url{https://cwe.mitre.org/top25/archive/2024/2024_cwe_top25.html}}.
\newblock
\newblock
\shownote{[Online; accessed 21-February-2025]}.


\bibitem[Neumann and Leis(2024)]%
        {NL24}
\bibfield{author}{\bibinfo{person}{Thomas Neumann} {and}
  \bibinfo{person}{Viktor Leis}.} \bibinfo{year}{2024}\natexlab{}.
\newblock \showarticletitle{A Critique of Modern {SQL} and a Proposal Towards a
  Simple and Expressive Query Language}. In \bibinfo{booktitle}{\emph{14th
  Conference on Innovative Data Systems Research, {CIDR} 2024, Chaminade, HI,
  USA, January 14-17, 2024}}.
\newblock


\bibitem[Nix and Dittrich(2025)]%
        {Nix2025ExtendingSQL}
\bibfield{author}{\bibinfo{person}{Joris Nix} {and} \bibinfo{person}{Jens
  Dittrich}.} \bibinfo{year}{2025}\natexlab{}.
\newblock \showarticletitle{Extending SQL to Return a Subdatabase}.
\newblock \bibinfo{journal}{\emph{{SIGMOD} '25}} (\bibinfo{year}{2025}).
\newblock


\bibitem[nplusone(2025)]%
        {djangonplusone}
\bibfield{author}{\bibinfo{person}{Django nplusone}.}
  \bibinfo{year}{2025}\natexlab{}.
\newblock \bibinfo{title}{{Django nplusone}}.
\newblock \bibinfo{howpublished}{\url{https://github.com/jmcarp/nplusone}}.
\newblock
\newblock
\shownote{[Online; accessed 13-March-2025]}.


\bibitem[Olston et~al\mbox{.}(2008)]%
        {piglatin}
\bibfield{author}{\bibinfo{person}{Christopher Olston},
  \bibinfo{person}{Benjamin Reed}, \bibinfo{person}{Utkarsh Srivastava},
  \bibinfo{person}{Ravi Kumar}, {and} \bibinfo{person}{Andrew Tomkins}.}
  \bibinfo{year}{2008}\natexlab{}.
\newblock \showarticletitle{Pig latin: a not-so-foreign language for data
  processing} \emph{(\bibinfo{series}{SIGMOD '08})}.
  \bibinfo{pages}{1099–1110}.
\newblock


\bibitem[Ong et~al\mbox{.}(2014)]%
        {ong_sql++}
\bibfield{author}{\bibinfo{person}{Kian~Win Ong}, \bibinfo{person}{Yannis
  Papakonstantinou}, {and} \bibinfo{person}{Romain Vernoux}.}
  \bibinfo{year}{2014}\natexlab{}.
\newblock \showarticletitle{The {SQL++} Semi-structured Data Model and Query
  Language: {A} Capabilities Survey of SQL-on-Hadoop, NoSQL and NewSQL
  Databases}.
\newblock \bibinfo{journal}{\emph{CoRR}}  \bibinfo{volume}{abs/1405.3631}
  (\bibinfo{year}{2014}).
\newblock
\showeprint[arXiv]{1405.3631}
\urldef\tempurl%
\url{http://arxiv.org/abs/1405.3631}
\showURL{%
\tempurl}


\bibitem[Oracle(2025)]%
        {oracle}
\bibfield{author}{\bibinfo{person}{Oracle}.} \bibinfo{year}{2025}\natexlab{}.
\newblock \bibinfo{title}{{JSON Relational Duality}}.
\newblock
  \bibinfo{howpublished}{\url{https://blogs.oracle.com/database/post/json-relational-duality-app-dev}}.
\newblock
\newblock
\shownote{[Online; accessed 24-March-2025]}.


\bibitem[Parent and Spaccapietra(1984)]%
        {ParentS84}
\bibfield{author}{\bibinfo{person}{Christine Parent} {and}
  \bibinfo{person}{Stefano Spaccapietra}.} \bibinfo{year}{1984}\natexlab{}.
\newblock \showarticletitle{An Entity-Relationship Algebra}. In
  \bibinfo{booktitle}{\emph{Proceedings of the First International Conference
  on Data Engineering, April 24-27, 1984, Los Angeles, California, {USA}}}.
  \bibinfo{publisher}{{IEEE} Computer Society}, \bibinfo{pages}{500--507}.
\newblock


\bibitem[PRQL(2025)]%
        {PRQL}
\bibfield{author}{\bibinfo{person}{PRQL}.} \bibinfo{year}{2025}\natexlab{}.
\newblock \bibinfo{title}{{PRQL}}.
\newblock \bibinfo{howpublished}{\url{https://github.com/PRQL/prql}}.
\newblock
\newblock
\shownote{[Online; accessed 1-April-2025]}.


\bibitem[Rel(2025)]%
        {rel}
\bibfield{author}{\bibinfo{person}{Rel}.} \bibinfo{year}{2025}\natexlab{}.
\newblock \bibinfo{title}{{Rel}}.
\newblock \bibinfo{howpublished}{\url{https://docs.relational.ai/rel}}.
\newblock
\newblock
\shownote{[Online; accessed 2-April-2025]}.


\bibitem[seal(2025)]%
        {djangoseal}
\bibfield{author}{\bibinfo{person}{Django seal}.}
  \bibinfo{year}{2025}\natexlab{}.
\newblock \bibinfo{title}{{Django seal}}.
\newblock
  \bibinfo{howpublished}{\url{https://github.com/charettes/django-seal}}.
\newblock
\newblock
\shownote{[Online; accessed 13-March-2025]}.


\bibitem[Shute et~al\mbox{.}(2024)]%
        {ShuteBBBDKLMMSWWY24}
\bibfield{author}{\bibinfo{person}{Jeff Shute}, \bibinfo{person}{Shannon
  Bales}, \bibinfo{person}{Matthew Brown}, \bibinfo{person}{Jean{-}Daniel
  Browne}, \bibinfo{person}{Brandon Dolphin}, \bibinfo{person}{Romit
  Kudtarkar}, \bibinfo{person}{Andrey Litvinov}, \bibinfo{person}{Jingchi Ma},
  \bibinfo{person}{John~D. Morcos}, \bibinfo{person}{Michael Shen},
  \bibinfo{person}{David Wilhite}, \bibinfo{person}{Xi Wu}, {and}
  \bibinfo{person}{Lulan Yu}.} \bibinfo{year}{2024}\natexlab{}.
\newblock \showarticletitle{{SQL} has problems. We can fix them: Pipe syntax in
  {SQL}}.
\newblock \bibinfo{journal}{\emph{Proc. {VLDB} Endow.}} \bibinfo{volume}{17},
  \bibinfo{number}{12} (\bibinfo{year}{2024}), \bibinfo{pages}{4051--4063}.
\newblock


\bibitem[Yannakakis(1981)]%
        {yannakakis_acyclic-database-schemes}
\bibfield{author}{\bibinfo{person}{Mihalis Yannakakis}.}
  \bibinfo{year}{1981}\natexlab{}.
\newblock \showarticletitle{Algorithms for Acyclic Database Schemes}. In
  \bibinfo{booktitle}{\emph{Very Large Data Bases, 7th International
  Conference, September 9-11, 1981, Cannes, France, Proceedings}}.
  \bibinfo{publisher}{{IEEE} Computer Society}, \bibinfo{pages}{82--94}.
\newblock


\bibitem[Zaharia et~al\mbox{.}(2012)]%
        {ZahariaCDDMMFSS12}
\bibfield{author}{\bibinfo{person}{Matei Zaharia}, \bibinfo{person}{Mosharaf
  Chowdhury}, \bibinfo{person}{Tathagata Das}, \bibinfo{person}{Ankur Dave},
  \bibinfo{person}{Justin Ma}, \bibinfo{person}{Murphy McCauly},
  \bibinfo{person}{Michael~J. Franklin}, \bibinfo{person}{Scott Shenker}, {and}
  \bibinfo{person}{Ion Stoica}.} \bibinfo{year}{2012}\natexlab{}.
\newblock \showarticletitle{Resilient Distributed Datasets: {A} Fault-Tolerant
  Abstraction for In-Memory Cluster Computing}. In
  \bibinfo{booktitle}{\emph{Proceedings of the 9th {USENIX} Symposium on
  Networked Systems Design and Implementation, {NSDI} 2012, San Jose, CA, USA,
  April 25-27, 2012}}, \bibfield{editor}{\bibinfo{person}{Steven~D. Gribble}
  {and} \bibinfo{person}{Dina Katabi}} (Eds.). \bibinfo{publisher}{{USENIX}
  Association}, \bibinfo{pages}{15--28}.
\newblock


\end{thebibliography}

\end{document}